\documentclass[prl,aps,twocolumn,showpacs,showkeys,amsmath,amssymb,reprint,floatfix]{revtex4-1}

\usepackage{braket}
\usepackage{amssymb,amsmath,amsfonts}
\usepackage{color}
\usepackage[colorlinks,linkcolor=blue,urlcolor=blue,citecolor=blue,anchorcolor=blue]{hyperref}
\usepackage{graphicx}
\usepackage{scalerel}
\DeclareMathOperator*{\varsum}{\scalerel*{\Sigma}{\sum}}
\newtheorem{protocol}{Protocol}
\newcommand{\Smono}[0]{\ensuremath{\bar{\mathcal{S}}}}
\newcommand{\Tmono}[0]{\ensuremath{\bar{\mathcal{T}}}}

\begin{document}

\title{Multiphoton Scattering Tomography with Coherent States}

\author{Tom\'as Ramos}
\email{t.ramos.delrio@gmail.com}
\affiliation{Instituto de F{\'\i}sica Fundamental IFF-CSIC, Calle Serrano 113b, Madrid 28006, Spain}
\author{Juan Jos{\'e} Garc{\'\i}a-Ripoll}
\affiliation{Instituto de F{\'\i}sica Fundamental IFF-CSIC, Calle Serrano 113b, Madrid 28006, Spain}
\date{\today}

\begin{abstract}
In this work we develop an experimental procedure to interrogate the single- and multiphoton scattering matrices of an unknown quantum system interacting with propagating photons. Our proposal requires coherent state laser or microwave inputs and homodyne detection at the scatterer's output, and provides simultaneous information about multiple ---elastic and inelastic--- segments of the scattering matrix. The method is resilient to detector noise and its errors can be made arbitrarily small by combining experiments at various laser powers. Finally, we show that the tomography of scattering has to be performed using pulsed lasers to efficiently gather information about the nonlinear processes in the scatterer. 
\end{abstract}

\maketitle

It is now possible to achieve strong and ultrastrong coupling between quantum emitters and propagating photons using superconducting qubits\ \cite{astafiev10,hoi11,hoi13,forndiaz16}, atoms \cite{tiecke14,goban15,Psolano17} or quantum dots\ \cite{arcari14} in photonic circuits, or even molecules in free space\ \cite{Hwang09}. This has motivated a stunning progress in the theory of single- and multiphoton scattering using wave functions\ \cite{shen05} and the Bethe ansatz\ \cite{shen07}, as well as input-output theory\ \cite{fan10,caneva15}, diagrammatic calculations\ \cite{pletyukhov12,laakso14,hurst2017}, and path integral formalism\ \cite{shi09,shi15}. Very recently, the theory has even covered the ultrastrong coupling regime\ \cite{shi17}. Experiments, however, cannot yet recover all the scattering information predicted by those studies, and are limited to comparing low-power coherent state transmission coefficients\ \cite{astafiev10,forndiaz16,pechal16}, cross-Kerr phases\ \cite{hoi13}, and antibunching\ \cite{hoi11}. We therefore need an ambitious framework for reconstructing the complete \mbox{one-,} two-, or ideally any multiphoton scattering matrix. Such framework would allow studying the elastic\ \cite{peropadre13}, and inelastic\ \cite{goldstein13} properties of quantum impurities in waveguides, quasi-particle spectroscopy\ \cite{kurcz14}, interactions\ \cite{gorshkov10} in quantum simulators, and even characterizing all-optical quantum processors.

\begin{figure}[t]
\center
\includegraphics[width=1\linewidth]{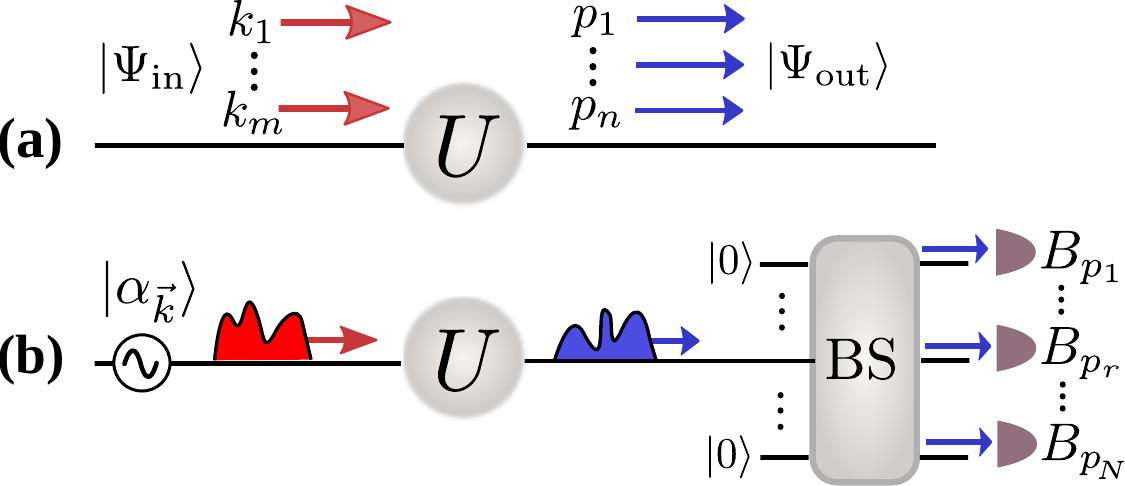}
\caption{(a) A quantum scatterer with scattering matrix $U$ transforms an input state $|\Psi_{\rm in}\rangle$ of $m$ photons with generic quantum numbers $k_1,\ldots,k_m$ into an outgoing state $|\Psi_{\rm out}\rangle=U\ket{\Psi_{\rm in}}$ of $n$ photons with labels, $p_1,\ldots, p_n$. (b) Our experimental protocol for determining $U$ requires coherent state wave packets inputs $|\alpha_{\vec{k}}\rangle$, prepared with a signal generator ($\sim$), and homodyne measurements at the output. Prior to measurement, the output signal is split evenly by an $N$-port beam splitter (${\rm BS}$), so as to measure all possible filtered correlations $\langle B_{p_1}\ldots B_{p_n}\rangle$.\label{fig:setup}}
\end{figure}

In this Letter we present a theoretical and experimental framework for estimating the scattering matrix
\begin{align}
S_{p_1\ldots p_n k_1\ldots k_m}=\braket{0|A_{p_1}\!\ldots A_{p_n} U A_{k_1}^\dag\! \ldots A_{k_m}^\dag|0},\label{matrixElements}
\end{align}
which describes the transition amplitude from an input state of $m$ photons with generic quantum numbers $k_1,\ldots, k_m$, to an asymptotic output state of $n$ photons with labels $p_1,\ldots, p_n$ [cf. Fig.\ \ref{fig:setup}a].
Operator $U$ represents the evolution in the limit of infinitely long time, and $A^\dag_k$ are generic input and output bosonic operators acting on the vacuum state $|0\rangle$. Our proposal assumes an experimental setup that injects coherent states and performs homodyne detection at the output of a generalized multiport beam splitter [cf. Fig.\ \ref{fig:setup}b]. Combining measurements with different input phases and amplitudes, we can approximate Eq.~(\ref{matrixElements}) with arbitrarily small error. The scheme is ideally suited for superconducting circuit and nanophotonic experiments, because all noise from amplifiers or detectors is canceled without previous calibration, similar to the dual-path method\ \cite{menzel10,dicandia14,dasilva10}.%
In this Letter we also prove that the nonlinear contributions in the scattering are efficiently activated only for finite length input wave packets $A_{k}^\dag$. However, standard deconvolution techniques applied to such experiments accurately and robustly reconstruct the scattering matrix in the monochromatic limit. We exemplify all these ideas for a two-level scatterer, whose scattering matrix is known\ \cite{fan10} and has only been probed at the single-photon level\ \cite{astafiev10,hoi11}. This Letter is closed with a discussion on the generality of the protocol, and possible experimental implementations.

\paragraph{Scattering tomography protocol.--}%
Our presentation begins with the tomography architecture, continues with a discrete set of input states and a corresponding set of measurements, and closes with a general reconstruction formula for the scattering matrix. As shown in Fig.~\ref{fig:setup}b, the setup consists of the quantum scatterer to be analyzed ---or any active or passive optical medium---, a photonic channel that couples the light in and out of the scatterer, a signal generator to prepare the input states, and a multiport beam splitter that divides the output signal into $N$ independent measurement ports.

\begin{figure}[t]
\center
\includegraphics[width=0.85\linewidth]{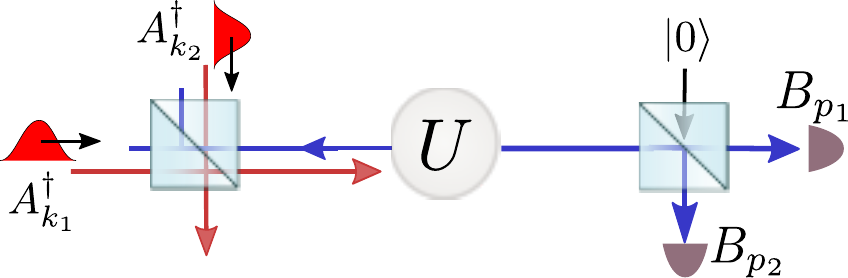}
\caption{A possible implementation of the protocol that works for one- and two-photon scattering matrices.}\label{fig:twoPhotonExp}
\end{figure}

Our protocol requires a specific set of input states to probe the scatterer: coherent state wave packet inputs $|\alpha_{\vec{k}}\rangle$, created on top of the vacuum \footnote{Throughout this work we assume that the scatterer has unique initial and final states, which are implicit in the definition of $\ket{0}$, but this restriction can be easily lifted.} through a superposition of $M$ bosonic modes $A_{k_j}^\dag$
\begin{equation}
\ket{\Psi_{\rm in}} = |\alpha_{\vec{k}}\rangle=e^{-\frac{1}{2}|\alpha|^2}\mathrm{exp}\Big(\sum_{j=1}^M\alpha_{k_j} A^\dag_{k_j}\Big)\ket{0},\label{generalCoherentInput}
\end{equation}
where $\alpha_{k_j}$ are complex weights and $|\alpha|^2$ the mean photon number \footnote{The normalization of the coherent state \eqref{generalCoherentInput} implies that the mean photon number is given by $|\alpha|^2=\sum_{j,j'=1}^M\alpha_{k_j}^\ast\alpha_{k_{j'}}f_{jj'}$, where the commutators $[A_{k_j},A_{k_{j'}}^\dag]=f_{jj'}\mathbb{I}$ are not orthonormal for wave packets.}. These multimode coherent states can be prepared using a signal generator [cf.~Fig.~\ref{fig:setup}b], or combining laser pulses through beam-splitters [cf.~Fig.~\ref{fig:twoPhotonExp}]. While later examples assume wave packets $A_{k_j}^\dag$ centered around momenta $k_j$, the formalism allows $k_j$ to label any set of quantum numbers: frequency, polarization, path, etc.

The output of the scatterer $|\Psi_{\rm out}\rangle=U|\Psi_{\rm in}\rangle$ is led through a balanced multiport beam splitter into $N$ homodyne detectors. At each output port, we filter outgoing photons with quantum numbers $\{p_r\}_{r=1}^N$, and measure the quadratures $\{X_{p_r},P_{p_r}\}$ to reconstruct the Fock operators $B_{p_r}=X_{p_r}+iP_{p_r}$. The nature of the beam splitter transformation is irrelevant, but all detectors should get a similar fraction of the scattered output, typically $B_{p_r}=N^{-1/2}A_{p_r}+\text{(N-1 vacuum inputs)}.$ Combining the homodyne measurements we estimate any correlation function of the form
\begin{align}
\langle B_{p_1}\ldots B_{p_n}\rangle=N^{-n/2}\langle \alpha_{\vec{k}}|U^\dag A_{p_1}\ldots A_{p_n} U|\alpha_{\vec{k}}\rangle,\label{generalQuadrature}
\end{align}
where the filtered labels $p_1,\ldots,p_n$ are to the outgoing indices in the scattering matrix sector to be estimated\ \eqref{matrixElements}. Note that any set size $n\leq N$ is possible.

The expectation value\ \eqref{generalQuadrature} is an analytic function $F_n[\vec{\alpha}]$ of the complex amplitudes $\alpha_{\vec{k}}$. Each order of the Taylor expansion of $F_n$ is determined by a different sector of the scattering matrix. As explained in the Supplemental Material \cite{SuppMat}, the values $F_n[\vec{\alpha}]$ for different complex amplitudes ---i.e. the correlation functions for different input states\ \eqref{generalCoherentInput}---, can be added together, so as to cancel all terms except a desired sector of the scattering matrix and a small error. This is the basis for our protocols.
\begin{protocol}[General]
Let us assume a setup such as the one in Fig.\ \ref{fig:setup}b. In order to reconstruct the scattering matrix\ \eqref{matrixElements} with $n\leq N$ and $m\leq M$, we will prepare $2^MM$ input states $\ket{\Psi_{\rm in}(l,\vec{s})}$, labeled by $l=1\ldots 2M$, and $\vec{s}=(s_1,\ldots,s_M)$. The input states~\eqref{generalCoherentInput} built from $j=1\ldots M$ wave packets only differ in the choice of phases
  \begin{equation}
    \alpha_{k_j}^{l, \vec{s}}= s_j  e^{i\phi_l} |\alpha_{k_j}|,\;\;\;\;
    \left\{
      \begin{array}{l}s_1=1 \\ s_{m\geq 2}=\pm 1 \end{array}\right.,\;\;\;\; \phi_l=\frac{\pi}{M}l.
  \end{equation}
For each input state, we measure the $2N$ amplitudes $B_{p_r}$ repeatedly, gathering statistics to reconstruct all correlations $F_n(l,\vec{s})=\braket{\Psi_{\rm in}(l,\vec{s})|U^\dag B_{p_1}\ldots B_{p_n}U|\Psi_{\rm in}(l,\vec{s})}$, for $n=1\ldots N$. The scattering matrix is approximated as
\begin{equation}
S_{p_1\ldots p_n k_1\ldots k_m}=\frac{N^{n/2}e^{|\alpha|^2}}{2^{M}M}\sum_{l=1}^{2M}\sum_{\vec{s}}\frac{F_n(l,\vec{s})}{\prod_{j=1}^m \alpha_{k_j}^{l,\vec{s}}}+\varepsilon_{m}^{(1)},\label{generalResult}
\end{equation}
with a controlled error scaling as $\varepsilon_m^{(1)}=\mathcal{O}(|\alpha|^2)$.
\end{protocol}
Note how the same set of measurements outcomes provides a \emph{simultaneous} reconstruction of \emph{all} scattering matrices from sizes $1\times 1$ up to $N\times M$.

\paragraph{Elastic scatterers.--}%
The reconstruction protocol simplifies when scattering conserves the total number of photons. This happens for emitters with one ground state and Jaynes-Cummings type interactions with $U(1)$ symmetry (no cyclic transitions). The scattering matrix\ \eqref{matrixElements} is exactly zero for $n\neq m$, quadrature measurements do not depend on global input phases and the total number of measurement setups reduces to $2^{M-1}$.
\begin{protocol}[Elastic scatterers]
When it is a priori known that the scatterer conserves the photon number, we follow the steps in Protocol 1, but reduce the choice of input states to $\ket{\Psi_{\rm in}(\vec{s})}$, where $\vec{s}=(s_1,\ldots,s_M)$ and
  \begin{align}
    &\alpha_{k_j}^{\vec{s}}= s_j  |\alpha_{k_j}|,\;\;\;\;\;
      \left\{
      \begin{array}{l}s_1=1\\ s_{j\geq 2}=\pm 1\end{array}\right.\label{elasticFormula},\\
    &F_n(\vec{s}) =\braket{\Psi_{\rm in}(\vec{s})|U^\dag B_{p_1}\ldots B_{p_n}U|\Psi_{\rm in}(\vec{s})},\notag\\
    &S_{p_1\ldots p_m k_1\ldots k_m}=\frac{N^{m/2}e^{|\alpha|^2}}{ 2^{M-1}}\sum_{\vec{s}}\frac{F_m(\vec{s})}{\prod_{j=1}^m \alpha_{k_j}^{\vec{s}}}+\varepsilon_{m}^{(1)},\notag
\end{align}
with an error bounded by $|\varepsilon_m^{(1)}|\leq\mathcal{O}(e^{|\alpha|^{2}}-1)$.
\end{protocol}

\paragraph{Examples.--}%
Experiments with superconducting or optical qubits at low power are well described by a RWA Hamiltonian\ \cite{fan10}, and we can apply Protocol 2. For a single photon we require only one input with arbitrary, but small, complex amplitude $\alpha_{k_1}$, and the measurement of one quadrature $B_{p_1}$, obtaining
\begin{align}
S_{p_1k_1}=\braket{0|A_{p_1}U A_{k_1}^\dagger|0}=e^{|\alpha|^2}\frac{\langle B_{p_1}\rangle}{\alpha_{k_1}}+\varepsilon_{1}^{(1)}.
\end{align}
This formula includes the limit of state-of-the-art experiments\ \cite{astafiev10,hoi11,forndiaz16}, where the transmission and reflection coefficients of single photons with momentum $k$ and fixed polarization, $S_{k,k}$ and $S_{-k,k}$, are recovered from the ratio between the input amplitude $\alpha_{k}$ of a monochromatic coherent beam, and the scattered amplitude $\braket{B_{\pm k}}$.

The reconstruction of the two-photon scattering matrix demands at least two measurement ports, $B_{p_1}$ and $B_{p_2}$, and a set of two input modes, $A_{k_1}^\dag$ and $A_{k_2}^\dag$. As shown in Fig.~\ref{fig:twoPhotonExp}, an experiment could combine two independent pulses through a beam splitter, and then direct the scattering output to two homodyne measurement devices for estimating the correlations $\braket{B_{p_1}B_{p_2}}$. For a RWA model, we reconstruct
\begin{align}
S_{p_1p_2k_1k_2}=e^{|\alpha|^2}\frac{\left[F_2(1,1)-F_2(1,-1)\right]}{|\alpha_{k_1}||\alpha_{k_2}|}+\varepsilon_{2}^{(1)},\label{TwoPhotonExplicit}
\end{align}
using only two different input phases. If we cannot ensure $U(1)$ symmetry because of inelastic channels \cite{goldstein13}, ultrastrong coupling \cite{shi17}, external driving on the scatterer \cite{bishop2009}, etc., we need 8 input states with varying global phase $\phi_l=(\pi/2)l$, and the general reconstruction formula \eqref{generalResult}. However, the same measurements provide us with estimates for $S_{p_1k_1}$, $S_{p_1k_1k_2}$, $S_{p_1p_2k_1}$, and $S_{p_1p_2k_1k_2}$.

\paragraph{Arbitrary reconstruction error.--}%
There are four sources of error in our reconstruction protocol: (i) quantum fluctuations in quadrature measurements, (ii) detector noise, (iii) imperfect input preparation, and (iv) approximation error. The first source of error scales as $\mathcal{O}(\mathcal{N}^{-1/2})$ and can be decreased arbitrarily by increasing the number of repetitions $\mathcal{N}$ of the experiment. By design, our protocol is intrinsically resilient to the second source of errors, because detector noise averages out when combining odd powers of quadratures from different detectors ---similar to Refs.\ \cite{menzel10,dicandia14,dasilva10}.

Imperfections in the preparation of relative phases $s_j=\pm1+\delta s_j^{(\pm)}$, laser power $|\alpha|^2+\delta n$, and global phases $\phi_l=\pi l/M+\delta \phi_l$, add linear contributions to the error, $\varepsilon_{\rm sign}\sim |\delta s_j^{(\pm)}|$, $\varepsilon_{\alpha}\sim|\delta n|/|\alpha|^2$ and  $\varepsilon_{\phi}\sim |\delta \phi_l|/|\alpha|^{m-1}$\cite{SuppMat}. These errors can be controlled using standard calibration and phase stability techniques, and ensuring to work at moderate powers, $|\alpha|^2\sim 1$.

\begin{figure}[t]
\center
\includegraphics[width=\linewidth]{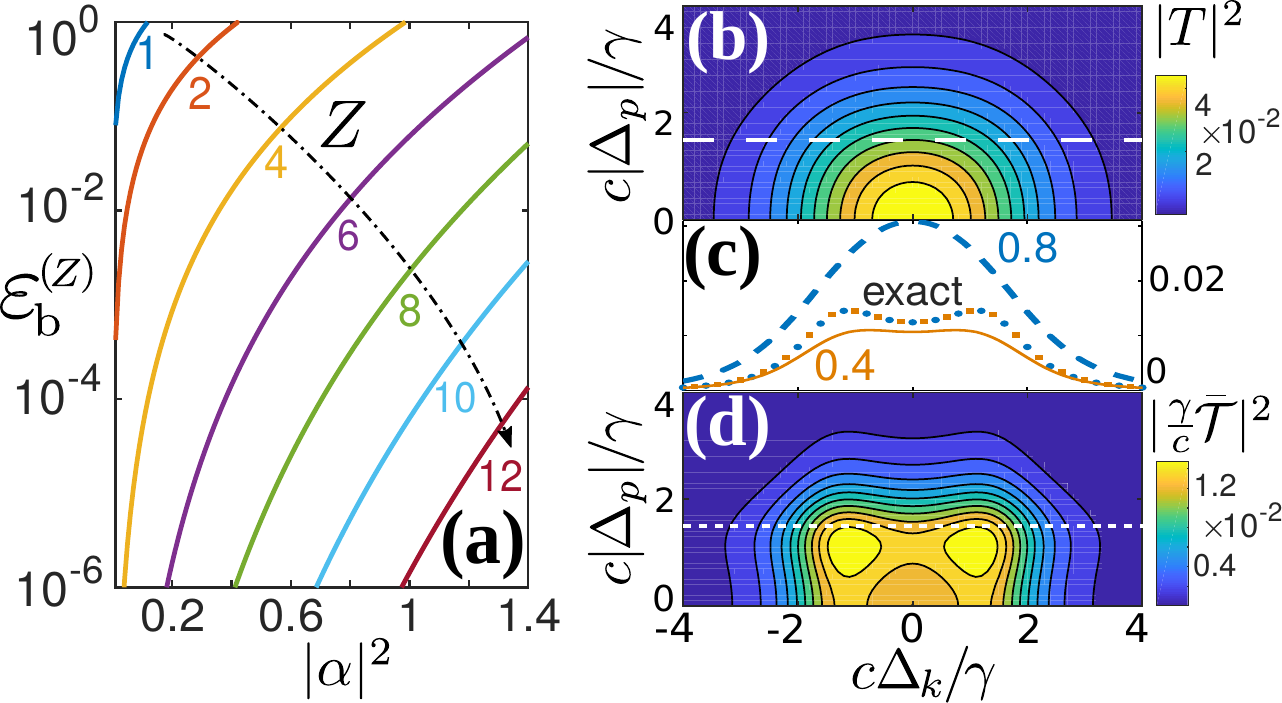}
\caption{Reconstruction of the nonlinear two-photon scattering matrix for a two-level system weakly coupled to a 1D photonic channel. (a) Upper bound of the reconstruction error $\varepsilon_{\rm b}^{(Z)}\geq|\varepsilon_2^{(Z)}|$, as derived in the Supplemental Material \cite{SuppMat} when combining $q=1,\ldots, Z$ scattering matrix estimates $E(b^{q-1}|\alpha|^2)$ with $b=1.05$. The various curves show $\varepsilon_{\rm b}^{(Z)}$ as a function of the smallest laser power $|\alpha|^2$ for $Z=1,2,\dots, 12$ (from top to bottom). (b) Predicted transmission measurement of $|T_{p_1p_2k_1k_2}|^2$ for a qubit with decay rate $\gamma$ and gaussian wave packets of width $\sigma=0.8\gamma/c$. We vary the momentum differences $\Delta_k=(k_2-k_1)/2$ and $\Delta_p=(p_2-p_1)/2$, between incoming and outgoing photons, and fix the conserved average momentum to $\hat{k}=(\omega_0+1.5\gamma)/c$, with $\omega_0$ the qubit's transition frequency. (c) Cross sections of $|T|^2$ measured at different widths, $\sigma=0.8\gamma/c$ (dashed blue) and $\sigma=0.4\gamma/c$ (solid brown), give the same exact result for $|\gamma\Tmono/c|^2$ after deconvolution (blue and brown dotted line). As marked in (a) and (d), the cross sections correspond to $|\Delta_p|=1.5\gamma/c$. (d) Deconvolution of the measurements according to Eq.~\eqref{inverseTwoPhoton}, to recover the two-photon interaction strength $|\Tmono_{p_1p_2k_1k_2}|^2$ of the two-level scatterer derived in Refs.~\cite{shen07,fan10,SuppMat}. \label{fig:deconvolution}}
\end{figure}

Indeed, a feature of our method is that we are not restricted to working at infinitesimal $|\alpha|^2$. Instead, we can combine different estimates of the scattering matrix, $E(|\alpha|^2)=S-\varepsilon_m^{(1)}$, reconstructed from Eqs.~\eqref{generalResult}-\eqref{elasticFormula} at different laser powers $|\alpha|^2$, to create a refined estimate with a higher order truncation error $\varepsilon_m^{(Z)}=\mathcal{O}(|\alpha|^{2Z})$. The simplest instance of this idea requires one extra estimate
\begin{align}
S=\frac{b E(|\alpha|^2)-E(b|\alpha|^2)}{b-1} + \varepsilon_m^{(2)},
\end{align}
at a larger power $b>1$, to give $\varepsilon_m^{(2)}=\mathcal{O}(|\alpha|^4)$. Higher order formulas can be derived analytically\ \cite{SuppMat}, with error estimates $\varepsilon^{(Z)}_m=\mathcal{O}(|\alpha|^{2Z})$ that are strongly suppressed and allow working at $|\alpha|^2\gtrsim 1$. This is illustrated by Fig.~\ref{fig:deconvolution}a, where we plot an upper bound for $\varepsilon^{(Z)}_m(|\alpha|^2)$ for elastic two-photon scattering, as a function of the smallest laser power $|\alpha|^2$, for different approximation orders $Z$. For $|\alpha|^2\sim 1$, we just need $Z\sim 10$ and $b=1.05$, to get error bounds $\varepsilon_{\rm b}^{(Z)}\sim 10^{-4}$, showing the potential of this method to estimate multiphoton scattering matrices.

\paragraph{The need of wave packets.--}%
We will now discuss the case in which $k_j$ describes the momentum degree of freedom. We will argue that our reconstruction protocol requires input states that are finite length wave packets
\begin{align}
A_{k_j}^\dag = \int\mathrm{d}k'\ \psi_{k_j}(k')a^\dag_{k'},\label{wavepacket}
\end{align}
built from normalized superpositions of plane waves $a_{k'}^\dag$, with $[a_k,a_{k'}^\dag]=\delta(k-k')$ and $\int|\psi_{k_j}(k')|^2\mathrm{d}k'=1$. The discussion below does not include other discrete degrees of freedom, which can be straightforwardly added\ \cite{SuppMat}.

The use of pulsed light contrasts with existing theory, which computes the scattering matrix elements for well defined momentum modes\ \cite{fan10,caneva15,shi15}, as in
\begin{equation}
  \Smono_{p_1\ldots p_nk_1\ldots k_m}=\braket{0|a_{p_1}\ldots a_{p_n}Ua_{k_1}^\dagger\ldots a_{k_m}^\dag|0}.\label{monoS}
\end{equation}
The reasons for studying the monochromatic $\Smono$ are (i) the possibility of analytical calculations and that (ii) it reveals the underlying nonlinearity of the scatterer. Take, for instance, the two-photon scattering matrix for a two-level system, which can be decomposed as \ \cite{fan10,xu13}
\begin{align}
  \label{scatteringStructure}%
  \Smono_{p_1p_2k_1k_2}={}& \Smono_{p_1k_1}\Smono_{p_2k_2}+\Smono_{p_1k_2}\Smono_{p_2k_1} \notag\\ +&ic \Tmono_{p_1p_2k_1k_2}\delta(\omega_{p_1}+\omega_{p_2}-\omega_{k_1}-\omega_{k_2}),
\end{align}
with $\omega_k$ the photon dispersion relation and $c$ the velocity of light. The first two terms in Eq.~\eqref{scatteringStructure} connect independent single-photon events $\Smono_{pk}$, while the last one is \emph{the} nonlinear contribution $\Tmono_{p_1p_2k_1k_2}$ that describes photon-photon interaction mediated by simultaneous interaction with the scatterer, such as the two-photon Kerr effect\ \cite{hoi13}.

Interestingly, $S$ and $\Smono$ are related through an integral,
\begin{align}
  \label{relationS}
  {S}_{p_1\ldots p_n,k_1\ldots k_m}=\int\!\ldots\!\int &\mathrm{d}^{m}k'\mathrm{d}^{n}p'\ \Smono_{p'_1\ldots p'_n,k'_1\ldots k'_m}\times\\&\times \prod_{j=1}^m \psi_{k_j}(k_j')\prod_{r=1}^n\psi_{p_r}^\ast(p_r').\notag
\end{align}
We will evaluate this for the two-photon scattering matrix in a tomography experiment using Gaussian pulses $\psi_k(k')= G_\sigma(k'-k)^{1/2}$, with
\begin{align}
G_\sigma(k')&=(\pi\sigma^2)^{-1/2}e^{-(k'/\sigma)^2}\label{gaussian}.
\end{align}
We are specially interested in analyzing how the nonlinearity $\Tmono$ manifests in the monochromatic limit of negligible bandwidth $\sigma\to0$. To do so, we focus on forward scattering $(k_j,p_r>0)$ in a waveguide with linear dispersion relation $\omega_k=c|k|$, but this is easily extended\ \cite{SuppMat}. The main result is that the \emph{measured} two-photon scattering matrix $S_{p_1p_2,k_1k_2}=S_{p_1k_1}S_{p_2k_2}+S_{p_1k_2}S_{p_2k_1}+iT_{p_1p_2k_1k_2}$ also splits into single- and two-photon contributions, which are given by the convolutions,
\begin{align}
&S_{p_1k_1}=e^{-\frac{(p_1-k_1)^2}{4\sigma^2}}\int \mathrm{d}k_1'G_\sigma(k_1'-\frac{[k_1+p_1]}{2})t_{k_1'},\label{convolutionSingle}\\
&T_{p_1p_2k_1k_2}=2\sigma\sqrt{\pi}e^{-(p_1+p_2-k_1-k_2)^2/(8\sigma^2)}\times\label{convolutionTwoPhoton}\\
{}&\qquad\qquad\quad\times \iiint \mathrm{d}p_1'\mathrm{d}k_1'\mathrm{d}k_2'\ \mathcal{W}(p_1',k_1',k_2') \Tmono_{p_1'p_2'k_1'k_2'}.\notag
\end{align}
The transmission coefficient $t_{k_1'}$ appears in Eq.~\eqref{convolutionSingle} as a consequence of energy conservation, $\Smono_{pk}=t_p\delta(p-k)$, as well as the relation $p_2'=k_1'+k_2'-p_1'$ for the integration momenta in Eq.~\eqref{convolutionTwoPhoton}. Because the kernel $\mathcal{W}$ in Eq.~\eqref{convolutionTwoPhoton} is a product of three Gaussians\ \cite{SuppMat} and $\Tmono$ is typically a smooth and bounded function, any nonlinear contribution to the scattering experiment vanishes as we make the wave packet width $\sigma$ tend to zero,
\begin{equation}
	S_{p_1k_1}=\mathcal{O}(1),\qquad T_{p_1p_2k_1k_2} = \mathcal{O}(\sigma^1).
\end{equation}
This argument can be extended to higher order processes, $T_{p_1\ldots p_n k_1\ldots k_m}\sim \sigma^{(m+n-2)/2}$\ \footnote{This derives from Eq.~\eqref{relationS} by a purely dimensional analysis, assuming $\psi_k(k')=\tilde{\psi}_k(k'/\sigma)/\sigma^{1/2}$, and a monochromatic nonlinear contribution $\Smono\sim\Tmono\delta(\sum_{r}\omega_{p_r}-\sum_{j}\omega_{k_j})$ that satisfies energy conservation as in Eq.~\eqref{scatteringStructure}.}, illustrating the fact that nonlinear terms can only be activated when photons coexist in the scatterer, and the probability of this overlap tends to zero as the wave packet length $1/\sigma$ tends to infinity. We, therefore, conclude that an \emph{efficient reconstruction of the full scattering matrix for two or more photons requires working with finite duration wave packets}.

\paragraph{Deconvolution formulas.--} %
Even if we need wave packets to get an experimentally measurable signal, we can still reconstruct the monochromatic properties from such experiments using standard deconvolution techniques \cite{ulmer2003}. We illustrate this by deriving the single- and two-photon forward scattering coefficients $t_{k_1}$ and $\Tmono_{p_1p_2k_1k_2}$ from the measured $S_{k_1k_1}$ and $T_{p_1p_2k_1k_2}$. This requires inverting Eqs.~\eqref{convolutionSingle}-\eqref{convolutionTwoPhoton}, which can be done analytically for Gaussian wave packets \cite{ulmer2003,ulmer2010,fang1994}. For the single-photon transmission we obtain,
\begin{align}
  &t_{k_1}=\int\mathrm{d}k_1'\ \mathcal{K}_{\sigma}(k_1'-k_1)S_{k_1'k_1'},\mbox{ where}\notag\\
  &\mathcal{K}_{\sigma}(k)=G_\sigma(k)\sum_{q=0}^\infty \frac{(-1)^q}{2^q q!}H_{2q}\left(\frac{k}{\sigma}\right).\label{inverseOnePhoton}
\end{align}
The inverse kernel $\mathcal{K}_{\sigma}(k)$ contains Hermite polynomials $H_{q}(x)=(-1)^q e^{x^2}\partial_x^q (e^{-x^2})$ and produces a convergent series provided the function to deconvolve is $L_1$ integrable \cite{ulmer2003,ulmer2010}. As discussed above, the single-photon reconstruction still works with monochromatic beams, recovering the state-of-the-art experimental formula $t_{k_1}=\lim_{\sigma\to 0}S_{k_1k_1}$.

The reconstruction of the two-photon scattering strength $\Tmono$ from the measured values of $T$ involves a three-dimensional deconvolution using a product of the same inverse kernels $\cal{K}_\sigma$ \cite{ulmer2010}. Energy conservation imposes that the only nonzero elements of $\Tmono_{p_1p_2k_1k_2}$ have to be functions of the conserved average momentum $\hat{k}=(k_1+k_2)/2=(p_1+p_2)/2$, and the relative differences, $\Delta_p=(p_2-p_1)/2$ and $\Delta_k=(k_2-k_1)/2$. For these elements we get,
\begin{align}
\label{inverseTwoPhoton}
&\Tmono_{p_1p_2k_1k_2}\!=\!\frac{1}{\sqrt{\pi}\sigma}\!\iiint\!\mathrm{d}\hat{k}'\mathrm{d}\Delta_p'\mathrm{d}\Delta_k' T_{\hat{k}'-\Delta_p',\hat{k}'+\Delta_p',\hat{k}'-\Delta_k',\hat{k}'+\Delta_k'}\notag\\
&\quad\times\mathcal{K}_{\sigma}(\sqrt{2}[\hat{k}'-\hat{k}])\mathcal{K}_{\sigma}(\Delta_p'-\Delta_p)\mathcal{K}_{\sigma}(\Delta_k'-\Delta_k).
\end{align}

As an illustration, we evaluate the measured scattering matrix $T$ and the reconstructed monochromatic version $\Tmono$, for a gedanken experiment with Gaussian pulses and a two-level scatterer. This problem admits an analytical solution\ \cite{shen07,fan10,SuppMat} with which we can test the reconstruction formulas (\ref{convolutionTwoPhoton}) and (\ref{inverseTwoPhoton}). As shown in Figs.\ \ref{fig:deconvolution}b and \ref{fig:deconvolution}d, the measured matrix $|T|^2$ is broader than the monochromatic $|\Tmono|^2$, due to the convolution with the Gaussians. However, we have found that provided the wave packet size remains on order of the scatterer linewidth $\sigma \sim\gamma/c$, we can efficiently reconstruct $\Tmono$ from $T$. We exemplify this in Fig.~\ref{fig:deconvolution}c, where we show how cross sections of $|T|^2$ obtained for two different widths $\sigma=0.4\gamma/c$ (brown) and $\sigma=0.8\gamma/c$ (blue) both reconstruct the exact result for $|\Tmono|^2$ after the deconvolution (brown and blue dots). The fact that $\sigma \sim \gamma/c$ is a good compromise should not be a surprise, as this is the regime which maximizes the nonlinear effects and the coexistence of photons. In general, the gaussian deconvolution with kernel (\ref{inverseOnePhoton}) allows us to efficiently reconstruct any smooth sector of the scattering matrix, provided the measured function to deconvolve is $L_1$ integrable \cite{ulmer2003,ulmer2010} and the order of the width $\sigma$ is chosen according to the `bandwidth' of the specific scatterer.

\paragraph{Summary and outlook.--}%
This Letter introduced a tomography protocol for reconstructing the scattering matrix of a photonic field interacting with a quantum scatterer, using coherent states and correlated homodyne measurements. We have demonstrated that pulsed spectroscopy is needed to gather information about the nonlinear processes in scattering. This could remind the reader of two-dimensional pulsed spectroscopy methods in the optical and NMR realms\ \cite{keusters99,cho08}, but those constitute a time-resolved interrogation of the scatterer, whereas our protocol studies the asymptotic transformation\ \eqref{matrixElements} imparted by an optical medium in a propagating field.

While our protocol is inspired by recent progress in the fields of waveguide QED and nanophotonics, the idea, setup, and formulas can be used to probe any system that is in contact with a linear bosonic field. This includes not only superconducting qubits in strong-coupling\ \cite{astafiev10,hoi13} or ultrastrong-coupling 1D setups\ \cite{forndiaz16}, but also studying single molecule emitters in three dimensions\ \cite{Hwang09}, or other extended optical media. The reconstruction protocol is so general that it does not require any a-priori knowledge of the quantum emitter, and can be applied in the presence of decoherence and dissipation. We believe that under such circumstances our protocol is optimal, but particular symmetries or a better understanding of the models can lead to substantial simplifications to be considered in future work.

\begin{acknowledgments}
The authors acknowledge support from the MINECO/FEDER Project FIS2015-70856-P and CAM PRICYT Research Network QUITEMAD+ S2013/ICE-2801.
\end{acknowledgments}

\bibliographystyle{apsrev4-1}
%

\newpage
\onecolumngrid
\newpage
{
  \center \bf \large 
  Supplemental Material for: \\
  Multiphoton Scattering Tomography with Coherent States
  \vspace*{0.1cm}\\ 
  \vspace*{0.0cm}
}
\begin{center}
  Tom\'as Ramos and Juan Jos{\'e} Garc{\'\i}a-Ripoll\\
  \vspace*{0.15cm}
  \small{\textit{Instituto de F{\'\i}sica Fundamental IFF-CSIC, Calle Serrano 113b, Madrid 28006, Spain}}\\
  \vspace*{0.25cm}
\end{center}
\twocolumngrid

\setcounter{page}{1}

\section*{Contents}
\begin{itemize}
\item I.---Derivation of the general scattering tomography relations.
\item II.---Arbitrary reconstruction error by combining multiple estimates of the scattering matrix.
\item III.---Deconvolution with gaussian wave packets in dispersive channels.
\item IV.---Imperfections in the input state preparation.
\end{itemize}

\section{I.~Scattering tomography relations}

In this section, we derive the central relations underlying our multiphoton scattering tomography protocol. We relate the components of an unknown unitary matrix $U$ with the measurement of certain photonic quadrature correlations at the output, when probing the system with coherent state inputs (see Fig.~\ref{fig:singlechannel}). In a first order approximation, a small reconstruction error requires attenuated coherent state inputs, but this restriction is lifted in Sec.~II by combining various estimates at different input powers.

The derivation is split into two parts. First, in Sec.~I.A, we show how to reconstruct the scattering matrix elements in the total photon number basis, namely
\begin{align}
S_{nm}=\langle 0|A^n U (A^\dag)^m |0 \rangle.\label{ScatElementsTotal}
\end{align}
These matrix elements describe processes involving $m$ incoming and $n$ outgoing photons, without resolving other photonic degrees of freedom. Then, in Sec.~I.B, we extend the protocol to identify photons by a generic quantum number $k$, which accounts for any combination of external and/or internal photonic properties such as momentum, frequency, polarization, etc. In particular, we can reconstruct general scattering matrix elements of the form,
\begin{align}
  S_{p_1\ldots p_n k_1\ldots k_m}=\langle 0|A_{p_1}\ldots A_{p_n} U A^\dag_{k_1}\ldots A^\dag_{k_m} |0 \rangle.\label{ScatElementsProperty}
\end{align}
describing processes from $m$ incoming photons with general quantum numbers $k_1,\ldots, k_m$ to $n$ outgoing with $p_1,\ldots, p_n$. As explained below, the injection and detection of photons with specific quantum numbers requires the use of a signal generator at the laser input and a multiport beam splitter at the measurement output (see also Fig.~\ref{fig:singlechannel}b).

\subsection{A.~Tomography of a scattering matrix in the photon number basis}\label{tomographyJ}

\begin{figure}[t]
\center
\includegraphics[width=0.5\textwidth]{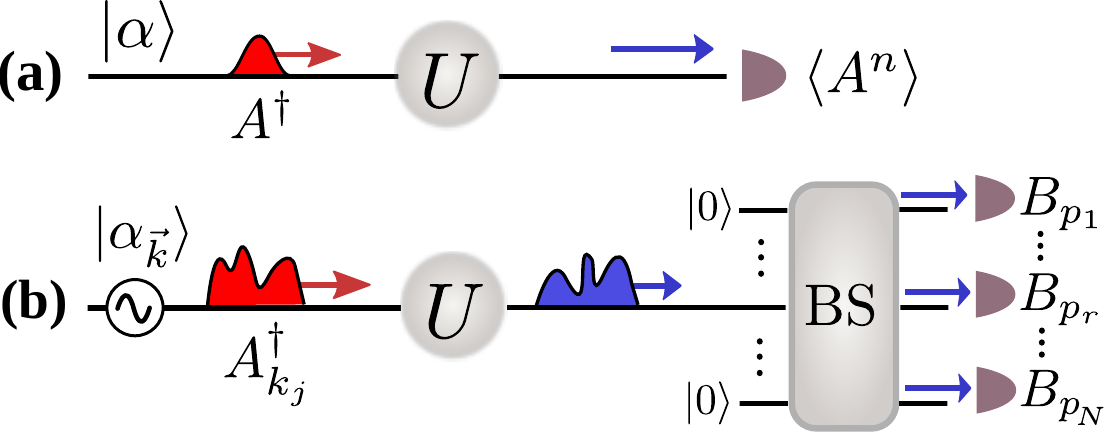}
\caption{Multiphoton scattering tomography protocol. (a) We input a coherent state $|\alpha\rangle$ and measure quadrature moments $\langle A^n\rangle$ at the output, to determine the scattering matrix elements in the total number basis \eqref{ScatElementsTotal}. (b) To have additional resolution on a generic photon quantum number $k$ \eqref{ScatElementsProperty}, we need to input a multimode coherent state $|\alpha_{\vec{k}}\rangle$ with different values of this quantity $\vec{k}=(k_1,\ldots,k_M)$, and measure the quadrature correlations $\langle B_{p_1}\ldots B_{p_n}\rangle$ at the $n=1,\ldots,N$ independent outputs of a beam splitter. \label{fig:singlechannel}}
\end{figure}

We start by considering the situation sketched in Fig.~\ref{fig:singlechannel}a, where photons propagate along a one-dimensional (1D) channel and interact with an unknown quantum medium according to a scattering matrix $U$. As a result, any photonic input state $|\Psi_{\rm in}\rangle$ is transformed into the output $|\Psi_{\rm out}\rangle=U|\Psi_{\rm in}\rangle$, with the only restriction that the vacuum state $|0\rangle$ remains invariant, i.~e.~$U|0\rangle=|0\rangle$. Besides this last condition, the unitary $U$ is completely arbitrary, and we show in the following how to reconstruct its elements using homodyne detection.

First, our scattering tomography protocol requires the preparation of a coherent state input,
\begin{align}
  |\Psi_{\rm in}\rangle=|\alpha\rangle=e^{-\frac{1}{2}|\alpha|^2}e^{\alpha A^\dag}|0\rangle,\label{coherentInputj}
\end{align}
where $A^\dag$ is a bosonic creation operator of a photon in the channel, satisfying $[A,A^\dag]=\mathbb{I}$, and $\alpha=|\alpha|e^{i\phi}$ is the complex vacuum displacement with module $|\alpha|$ and phase $\phi$.

Secondly, we require the measurement of photon output correlations of the form
\begin{align}
\langle A^n \rangle=\langle \Psi_{\rm out}|A^n|\Psi_{\rm out}\rangle=\langle \alpha|U^\dag A^n U|\alpha \rangle,\label{BnCorr}
\end{align}
which can be determined to any order $n$ by measuring the output quadratures $X$ and $P$ via homodyne detection, with $A=X+iP$. 

To derive the relation between the scattering matrix elements \eqref{ScatElementsTotal} and the correlations \eqref{BnCorr}, let us first consider the measurement of a general operator $Q$ at the output, namely  $\langle Q\rangle=\langle \alpha|U^\dag Q U|\alpha\rangle$. If we replace here the definition of the coherent state \eqref{coherentInputj} and decompose $\alpha$ in terms of its module and phase, we obtain
\begin{align}
  \langle Q\rangle=e^{-|\alpha|^2}\sum_{t=0}^{\infty}\sum_{r=\lambda}^{\infty}|\alpha|^{r+t} e^{i\phi(r-t)} C_{tr}.\label{systemOfequations}
\end{align}
where the coefficients $C_{tr}$ read, 
\begin{align}
C_{tr}=\frac{\langle 0|A^t U^\dag Q U (A^\dag)^r|0\rangle}{t!r!}.\label{Ccoeffs}
\end{align}
In addition, $\lambda$ can only take two values: $\lambda=1$ in the special case $Q|0\rangle=0$ (since then $C_{t0}=0$), and $\lambda=0$ in any other case. In particular, for the choice $Q=A^n$ we can simply take $\lambda=1$, but we keep the derivation general. 

Our aim now is to solve for the coefficients of the form $C_{0m}=\langle 0|U^\dag Q U (A^\dag)^m |0\rangle/m!$ from Eq.~\eqref{systemOfequations}, since they correspond to the scattering matrix elements \eqref{ScatElementsTotal} when $Q=A^n$. As we show in the following, we can isolate the coefficients $C_{0m}$, for $m=\lambda,\ldots, M$, by preparing $R=2(M+1-\lambda)$ different coherent input states $|\alpha(l)\rangle$ of the form \eqref{coherentInputj}, and measuring $\langle Q \rangle(l)=\langle \alpha(l)|U^\dag Q U|\alpha(l)\rangle$ for each of them ($l=1,\ldots, R$). Importantly, we choose the different coherent state inputs $|\alpha(l)\rangle$ to have the same module $|\alpha|$, but different global phase $\phi_l$ as $\alpha(l)=|\alpha|e^{i\phi_l}$. As a result, we can replace $\langle Q\rangle\rightarrow \langle Q \rangle(l)$ and $\phi\rightarrow\phi_l$ in Eq.~\eqref{systemOfequations}, obtaining a set of $l=1,\ldots, R$ independent equations for the unknowns $C_{tr}$, which read
\begin{align}
\langle Q \rangle(l)=e^{-|\alpha|^2}\sum_{t=0}^{\infty}\sum_{r=\lambda}^{\infty}|\alpha|^{r+t} e^{i\phi_l(r-t)} C_{tr}.\label{systemOfequations2}
\end{align}

To see how to choose the phases $\phi_l$ in order to solve for the coefficients $C_{0m}$ in Eq.~\eqref{systemOfequations2}, let us multiply Eq.~\eqref{systemOfequations2} by $e^{|\alpha|^2}e^{-im\phi_l}/R$ and sum over all values $l=1,\ldots, R$, obtaining the equation,
\begin{align}
\frac{e^{|\alpha|^2}}{R}\!\sum_{l=1}^{R}\frac{\langle Q \rangle(l)}{e^{im\phi_l}}\!=\!\sum_{u=\lambda}^{\infty}\varsum_{d=-u+2\lambda}^{u}\!\!\!\!|\alpha|^u T_{ud}\sum_{l=1}^{R}\frac{e^{i\phi_l(d-m)}}{R}\!,\label{intermediateJ}
\end{align}
after the convenient change of variables, $u=r+t$ and $d=r-t$. We have defined the new coefficients,
\begin{align}
T_{ud}=C_{(u-d)/2,(u+d)/2},\label{TCoeffs}
\end{align}
and the sum over $d$, denoted with a bold sum symbol $\varsum_{y=a}^b$, corresponds to a sum where $y$ goes from $a$ to $b$ in steps of $2$. Choosing the phases as 
\begin{align}
  \phi_l=(2\pi l)/R,\label{phaseschoice}
\end{align}
we see that the last factor on the right hand side of Eq.~\eqref{intermediateJ} becomes the discrete Fourier transform of a periodic Kronecker delta,
\begin{align}
\frac{1}{R}\sum_{l=1}^{R}e^{(2i\pi l/R)(d-m)}=\delta_{d,m+q R},\label{FourierDelta}
\end{align}
with $q$ taking any integer value due to the function periodicity of size $R$. When replacing \eqref{FourierDelta} into Eq.~\eqref{intermediateJ}, we see that all non-vanishing terms must satisfy $d=m+qR$, with $q\in \mathbb{Z}$ and $d$ of the same parity as $u$. This allows us to select specific terms in the sum. In particular, if we set 
\begin{align}
R=2(M+1-\lambda),\label{phasesMeasurement}
\end{align}
then all non-zero terms with $u\leq M+1$ in Eq.~\eqref{intermediateJ} can only take $d=m$ or equivalently $q=0$, as they all lie within the first period of the delta function. In addition, since $d$ is bounded by $-u+2\lambda\leq d \leq u$, then all terms with $u<m\leq M$ vanish, and our target coefficients $C_{0m}$ appear to lowest non-zero order $\sim |\alpha|^m$, for $m=\lambda,\ldots,M$, namely
\begin{align}
\frac{e^{|\alpha|^2}}{R}\!\sum_{l=1}^{R}\frac{\langle Q \rangle(l)}{e^{im\phi_j}}{}={}&|\alpha|^m C_{0m}\!+\!\!\!\varsum_{u=m+2}^{\infty}\!\!|\alpha|^u\!\!\!\sum_{q=q_{\rm min}}^{q_{\rm max}}\!\!\!T_{u,m+qR}\label{intermediateJ2}.
\end{align}
Here, the bounds of $q$ are given by $q_{\rm min}=-I[(u+m-2\lambda)/R]$ and $q_{\rm max}=I[(u-m)/R]$ where the function $I[x]$ rounds the number $x$ to its nearest smaller integer. From Eq.~\eqref{intermediateJ2} it is clear that in the case of attenuated coherent state inputs, $|\alpha|\ll 1$, all terms on order $\sim |\alpha|^{m+2}$ or higher are further suppressed and we can solve for $C_{0m}$ with a small relative error. By formally solving for the coefficient $C_{0m}$ in Eq.~\eqref{intermediateJ2}, we obtain the main result of this subsection,
\begin{align}
C_{0m}=\frac{\langle 0|QU(A^\dag)^m|0\rangle}{m!}{}&=\frac{\hspace{0.5mm}e^{|\alpha|^2}}{R}\!\sum_{l=1}^{R}\frac{\langle Q \rangle(l)}{[\alpha(l)]^m}+\epsilon_m^{(1)},\label{finalSingle}
\end{align}
which relates the desired matrix elements $\langle 0|QU(A^\dag)^m|0\rangle$, for powers $m=\lambda,\ldots, M$, with the measurements $\langle Q \rangle(l)=\langle \alpha(l)|U^\dag Q U|\alpha(l)\rangle$ at the output. In particular, when replacing $Q=A^n$ in Eq.~\eqref{finalSingle}, we access all the scattering matrix elements $S_{nm}=\langle 0|A^n U (A^\dag)^m|0\rangle$ in Eq.~\eqref{ScatElementsTotal}, describing $m=\lambda,\ldots, M$ incident photons to any number $n$ of outgoing photons in the channel $A$. As $Q|0\rangle=A^n|0\rangle=0$, we can take $\lambda=1$ and the number of measurements needed is $R=2M$.

Regarding the error $\epsilon_m^{(1)}$, we find a precise expression for it in terms of a series with even powers of the coherent state amplitude $|\alpha|$ as,
\begin{align}
\epsilon_m^{(1)}&=\sum_{v=1}^{\infty}|\alpha|^{2v}h_{vm}.\label{errorFirst}
\end{align}
From here it is clear that $\epsilon_m^{(1)}$ scales as $|\epsilon_m^{(1)}|\sim|\alpha|^{2}\ll 1$ in the case of attenuated coherent states. The coefficients $h_{vm}$ are independent of the coherent input power, and are given in general by,
\begin{align}
h_{vm}=-\sum_{q=q_{\rm min}}^{q_{\rm max}}C_{v-q\frac{R}{2},v+m+q\frac{R}{2}}.\label{errorCoefficients}
\end{align}
In addition, using Eqs.~\eqref{Ccoeffs}, \eqref{errorFirst} and \eqref{errorCoefficients}, as well as the triangular inequality, we can show that the error \eqref{errorFirst} is strictly bounded by $|\epsilon_m^{(1)}|\leq\!\sum_{v=1}^{\infty}|\alpha|^{2v}|h_{vm}|$, where the bound for the coefficients reads
\begin{align}
|h_{vm}|\leq\!\sum_{q=q_{\rm min}}^{q_{\rm max}}\!\!\!\frac{|\langle v-qR/2|U^\dag Q U|v+m+qR/2\rangle|}{\sqrt{(v+m+qR/2)!(v-qR/2)!}},\label{generalbound}
\end{align}
and $|v\rangle=(A^\dag)^{v}/\sqrt{v!}$ are the photon number states.

It is worth highlighting the special case where we \emph{a priori} know that the number of photons is conserved during the scattering, i.~e.~$[U,A^\dag A]=0$, since then the tomography protocol is strongly simplified. In particular, only the diagonal components $S_{mm}$ are nonzero and a single measurement, $R=1$, is enough to determine them. To see this, notice that the conservation of photon number and $Q=A^m$ implies that $T_{ud}$ is only nonzero for $d=m$. Using this in Eq.~\eqref{intermediateJ} allows us to derive the same Eqs.~\eqref{intermediateJ2} and \eqref{finalSingle}, but now valid for $R=1$ and all $m=1,\ldots,\infty$, since in this case all coefficients appearing at lower order than $\sim |\alpha|^m$ are zero regardless the value of $R$. Another peculiarity of this elastic scattering case is that the bound for the error in Eq.~\eqref{generalbound} takes a very simple closed form. Again, using $Q=A^m$ and the conservation of photon number, we can set $U|v\rangle=e^{i\varphi_v}|v\rangle$ and $q=0$ in Eq.~\eqref{generalbound}, and additionally using $A^m|v\rangle=\sqrt{v!/(v-m)!}|v-m\rangle$, we find a simple bound for the coefficients in this elastic case,
\begin{align}
|h_{vm}|\leq 1/v!.\label{boundh}  
\end{align}
As a result, the error $\epsilon_m$ in Eq.~\eqref{errorFirst} is strictly bounded by a displaced exponential for all $m$, namely
\begin{align}
|\epsilon_m^{(1)}|\leq \sum_{v=1}^{\infty}\frac{|\alpha|^{2v}}{v!}=e^{|\alpha|^2}-1.\label{strictbound}
\end{align}

\subsection{B.~Tomography of a scattering matrix with resolution on photon degrees of freedom}\label{tomographyL}

In this subsection, we extend the protocol to measure scattering matrix elements with resolution on a generic quantum number $k$, describing any combination of external or internal photon degrees of freedom as in Eq.~\eqref{ScatElementsProperty}. First, in Sec.~I.B.1.~, we generalize Eq.~\eqref{finalSingle} to determine matrix elements of the form $\langle 0|QU A^\dag_{k_1}\ldots A^\dag_{k_m}|0\rangle$, which describe input states of $m$ photons with different generic quantum numbers $k_1,\ldots, k_m$. Then, in Sec.~I.B.2.~, we give details on a beam splitter setup to measure the output quadrature correlations $\langle B_{p_1}\ldots B_{p_n}\rangle$ (see Fig.~\ref{fig:singlechannel}b), and we show how to reconstruct the general scattering matrix elements \eqref{ScatElementsProperty} from these correlations.

\subsubsection{1.~Multimode coherent state input}

First, we need a mechanism to inject photons, with various values of a combined quantum number $k$, at the input of the scatterer (see Fig.~\ref{fig:singlechannel}b). To this end we prepare a coherent state input $|\alpha_{\vec{k}}\rangle$ as in Eq.~\eqref{coherentInputj}, but now for a multimode superposition $A^\dag_{\vec{k}}$, defined for $M$ different components $\vec{k}=(k_1,\ldots,k_M)$ as 
\begin{align}
A^\dag_{\vec{k}}=\sum_{j=1}^M \xi_{j}A_{k_{j}}^\dag.\label{superpositionOp}
\end{align}
Here, $A_{k_j}^\dag$ for $j=1,\ldots, M$, are the creation operators of a photon with generic quantum number $k_j$, and $\xi_j$ is a projection vector normalized as $\sum_{j,v=1}^M \xi_{j}^\ast \xi_{v}[A_{k_j},A^\dag_{k_{v}}]=\mathbb{I}$, such that $[A_{\vec{k}},A_{\vec{k}}^\dag]=\mathbb{I}$. Notice that the modes $A_{k_j}$ do not satisfy standard commutation relations in general, allowing us to describe photon wave packets as inputs. For instance, for Gaussian wave packet modes as defined in Eqs.~\eqref{wavepacket} and \eqref{gaussian} of the main text, we obtain $[A_{k_j},A^\dag_{k_v}]=e^{-(k_j-k_{v})^2/(2\sigma)^2}\mathbb{I}$.

The coherent state input corresponding to the multimode superposition \eqref{superpositionOp} can be expressed more explicitly as
\begin{align}
  \ket{\Psi_{\rm in}}&= |\alpha_{\vec{k}}\rangle=e^{-\frac{1}{2}|\alpha|^2}e^{\alpha A^\dag_{\vec{k}}}\ket{0}\label{SuperInput0}\\
                     &=e^{-\frac{1}{2}|\alpha|^2}{\rm exp}\left({\sum_{j=1}^M\alpha_{k_j} A^\dag_{k_j}}\right)\ket{0},\label{SuperInput}
\end{align}
where the weights on each component are given by $\alpha_{k_j}=\alpha \xi_{j}$, with $\alpha=|\alpha|e^{i\phi}$, and the mean photon number reads $|\alpha|^2=\sum_{j,v=1}^M \alpha_{k_j}^\ast\alpha_{k_v}[A_{k_j},A^\dag_{k_v}]$. This multimode coherent state can be prepared using a signal generator (see Fig.~\ref{fig:singlechannel}b), which is readily implemented in microwave and optical photonic setups.

When we vary the global phase of the displacement weights as $\alpha_{k_j}(l)=|\alpha|e^{i\phi_l}\xi_j$, with $l=1,\ldots, R$ and $\phi_l$ defined in the previous subsection, we can apply the result \eqref{finalSingle} to the multimode superposition operator in Eq.~\eqref{superpositionOp}, and thereby determine the matrix elements $\langle 0|QU(A_{\vec{k}}^\dag)^m|0\rangle$, with the associated error $\epsilon_m^{(1)}=|\alpha|^2$. However, this is not enough the determine matrix elements of the form $\langle 0|QUA^\dag_{k_1}\ldots A^\dag_{k_m}|0\rangle$, since when expanding the multinomial in $\langle 0|QU(A_{\vec{k}}^\dag)^m|0\rangle$, we see that it contains many unwanted terms, namely
\begin{align}
  \langle 0|Q U(A_{\vec{k}}^\dag)^m|0\rangle=\!\!\!&\sum_{\substack{n_1,\ldots, n_M=1\\ [\sum n_j=m]}}^M\!\frac{m!(\xi_{1})^{n_1}\ldots (\xi_{M})^{n_M}}{n_1!\ldots n_M!}\label{multinomialDecom}\\ 
  {}&\hspace{0.6cm}\times\!\langle 0|QU(A^\dag_{k_1})^{n_1}\!\!\ldots (A^\dag_{k_M})^{n_M}|0\rangle.\notag
\end{align} 
Notice that the sums over $n_1,\ldots,n_M$ are constrained to values that satisfy $\sum_{j=1}^M n_j=m$, as demanded by the multinomial theorem. 

Our aim in this subsection is therefore to extract from the above relation the term $\langle 0|QUA^\dag_{k_1}\ldots A^\dag_{k_m}|0\rangle$ and relate it to $\langle 0|QU(A_{\vec{k}}^\dag)^m|0\rangle$ by canceling all unwanted terms. We achieve this by performing additional measurements as in the previous subsection, but now we vary the relative phases in the multimode superposition $A_{\vec{k}}^\dag(\vec{s})=\sum_{j=1}^M \xi_j^{(s_j)}A^\dag_{k_j}$ as $\xi_{j}^{(s_j)}=e^{i\Theta_{j}^{(s_j)}}|\xi_{j}|$. Here, $\Theta_{j}^{(s_j)}$ for $s_j=1,\ldots R_j$ denote the $R_j$ different values that the phase of each component $j=1,\ldots, M$ can take, and we keep all the moduli $|\xi_{j}|$ fixed. We also use the shorthand notation, $\vec{s}=(s_1,\ldots, s_M)$, to arrange the values of the $M$ different indices. To see how to choose the phases $\Theta_{j}^{(s_j)}$ in order to cancel the unwanted terms in Eq.~\eqref{multinomialDecom}, we divide this equation by $m!\ \xi_1^{(s_1)}\ldots \xi_m^{(s_m)}$ and sum over all independent values of $\vec{s}$, for $s_j=1,\ldots R_j$, and $j=1,\ldots, M$, obtaining
\begin{widetext}
\begin{align}
\sum_{s_1=1}^{R_1}\!\!\ldots\!\!\sum_{s_M=1}^{R_M}\frac{\langle 0|QU[ A^\dag_{\vec{k}}(\vec{s})]^m|0\rangle}{m!\xi_1^{(s_1)}\!\!\ldots \xi_m^{(s_m)}}=\!\!\!\!\sum_{\substack{n_1,\ldots, n_M=1\\ [\sum_j n_j=m]}}^M\!\!\!\!\!\!\frac{\langle 0|QU(A^\dag_{k_1})^{n_1}\!\!\ldots (A^\dag_{k_M})^{n_M}|0\rangle}{n_1!\ldots n_M!}\prod_{j=1}^{m}|\xi_{j}|^{n_j-1}\!\sum_{s_j=1}^{R_j}e^{i\Theta_j^{(s_j)}(n_j-1)}\!\!\!\!\prod_{v=m+1}^{M}\!\!|\xi_{v}|^{n_v}\!\!\sum_{s_v=1}^{R_v}\!e^{i\Theta_v^{(s_v)}n_v}.\label{superformula}
\end{align}
\end{widetext}
Choosing $\Theta_j^{(s_j)}=2\pi (s_j-1)/R_j$, for all $j=1,\ldots,M$, we can use the same property \eqref{FourierDelta} as in the previous subsection and form periodic Kronecker deltas in Eq.~\eqref{superformula},
\begin{align}
\sum_{s_j=1}^{R_j}\!e^{i\Theta_j^{(s_j)}(n_j-t_j)}=R_j \delta_{n_j,t_j+q_j R_j},
\end{align}
with $q_j$ and $t_j$ arbitrary integers. Importantly, due to the constrains $n_j\geq 0$ and $\sum_j n_j=m$, it is enough to choose $R_j$ as 
\begin{align}
R_{j\geq 2}=2,\qquad {\rm and}\qquad R_1=1,\label{RjChoice}
\end{align}
so that the deltas in Eq.~\eqref{superformula} cancel all terms, except for the one with $n_1=\ldots=n_m=1$ and $n_{m+1}=\ldots=n_R=0$, and we obtain the desired relation,
\begin{align}
\langle 0|QUA^\dag_{k_1}\!\ldots \!A^\dag_{k_m}|0\rangle=\!\!\!\sum_{s_2,\ldots,s_M=1}^{2}\frac{\langle 0|QU[A^\dag_{\vec{k}}{}(\vec{s})]^m|0\rangle}{\xi_1^{(1)}\xi_2^{(s_2)}\!\!\ldots \xi_m^{(s_m)}m!2^{M-1}}.\label{IntermediateL}
\end{align}

With the choice \eqref{RjChoice}, we see that the relative phases take the simple values $\Theta_{j\geq 2}^{(1,2)}=(0,\pi)$ and $\Theta_1^{(1)}=0$, resulting only in sign changes of the vector components as, $\xi_{j\geq 2}^{(1,2)}=\pm |\xi_j|$ and $\xi_1^{(1)}=|\xi_1|$. Therefore, from now on (and also in the main text), we conveniently redefine the indices $\vec{s}=(s_1,\ldots,s_M)$ as
\begin{align}
s_{j\geq 2}=\pm 1,\qquad {\rm and}\qquad s_1=1,\label{notationLabel}
\end{align}
such that the different values of vector components are simply given by
\begin{align}
\xi_j^{(s_j)}=s_j|\xi_j|.
\end{align}

Finally, replacing the result \eqref{finalSingle} into Eq.~\eqref{IntermediateL}, we obtain a closed relation for the matrix elements $\langle 0|QUA^\dag_{k_1}\ldots A^\dag_{k_m}|0\rangle$, with $m=\lambda,\ldots, M$, and the various measurements $\langle Q \rangle(l,\vec{s})=\langle \alpha_{\vec{k}}(l,\vec{s})|U^\dag Q U|\alpha_{\vec{k}}(l,\vec{s})\rangle$, given by
\begin{align}
{}&\langle 0|QUA^\dag_{k_1}\ldots A^\dag_{k_m}|0\rangle=\frac{e^{|\alpha|^2}}{2^{M-1}R}\sum_{l=1}^{R}\!\sum_{\vec{s}}\frac{\langle Q \rangle(l,\vec{s})}{\prod_{j=1}^m\alpha_{k_j}^{l,\vec{s}}}+\varepsilon_m^{(1)}.\label{resultProperty}
\end{align}
Here, we have used the new notation \eqref{notationLabel} for $\vec{s}$, and defined the weights $\alpha_{k_j}^{l,\vec{s}}$ of the coherent states $|\alpha_{\vec{k}}(l,\vec{s})\rangle$ in Eq.~\eqref{SuperInput} as  
\begin{align}
\alpha_{k_j}^{l,\vec{s}}=\alpha(l)\xi_j^{(s_j)}=s_je^{i\phi_l}|\xi_{j}||\alpha|=s_je^{i\phi_l}|\alpha_{k_j}|,
\end{align}
with $l=1,\ldots, R$, and $\phi_l$ as previously defined in Eqs.~\eqref{phaseschoice} and \eqref{phasesMeasurement}.

The error in Eq.~\eqref{resultProperty} is also of the form,
\begin{align}
\varepsilon_m^{(1)}=\sum_{v=1}^{\infty}|\alpha|^{2v}f_{vm},\label{secondError}
\end{align}
where the modified error coefficients $f_{vm}^{(1)}$ are given by
\begin{align}
f_{vm}=\frac{1}{2^{M-1}}\sum_{\vec{s}}\frac{h_{vm}(\vec{s})}{\prod_{j=1}^m\xi_j^{(s_j)}},\label{secondErrorCoeffs}
\end{align}
with $h_{vm}(\vec{s})$, defined in Eq.~\eqref{errorCoefficients}, now depends on $\vec{s}$ via $A_{\vec{k}}(\vec{s})$ in the coefficients \eqref{Ccoeffs}.

\subsubsection{2.~Homodyne detection at beam splitter outputs}\label{tomographyW} 

Notice that Eq.~\eqref{resultProperty} directly gives the desired scattering matrix elements,
\begin{align}
S_{p_1\ldots p_n k_1\ldots k_m}=\langle 0| A_{p_1}\ldots A_{p_n}UA_{k_1}^\dag\ldots A_{k_m}^\dag |0\rangle,
\end{align}
in the case we replace $Q=A_{p_1}\ldots A_{p_n}$.  Using the resulting relation, however, demands the measurement of specific correlation functions at the scatterer's output on channel $A$, namely
\begin{align}
\langle A_{p_1}\ldots A_{p_n} \rangle=\langle \alpha_{\vec{k}}|U^\dag A_{p_1}\ldots A_{p_n}U|\alpha_{\vec{k}}\rangle.\label{AACorr}
\end{align}

To distinguish the contribution of photons with different generic quantum numbers $p_1,\ldots, p_n$ in the above correlations \eqref{AACorr}, a filtering mechanism at the scatterer's output is needed. Therefore, we connect the output of the scatterer to a multiport beam splitter, which divides the scattered signal into $N$ independent channels (see Fig.~\ref{fig:singlechannel}b). At each independent output port of the beam splitter, labeled by $r=1,\ldots, N$, we measure the quadratures $X_{p_r}$ and $P_{p_r}$, and reconstruct the field operator $B_{p_r}=X_{p_r}+iP_{p_r}$ for outgoing photons with quantum number $p_r$. When gathering enough statistics via homodyne detection, we can determine the correlations,
\begin{align}
\langle B_{p_1}\ldots B_{p_n}\rangle=\langle \alpha_{\vec{k}}|U^\dag B_{p_1}\ldots B_{p_n} U|\alpha_{\vec{k}}\rangle,\label{BcorrsFinall}
\end{align}
which can be related to the correlations in Eq.~\eqref{AACorr} via the beam splitter transformation $U_{\rm BS}$.

Notice that the specific form of the beam splitter transformation $U_{\rm BS}$ is irrelevant, as long as each output port $r$ gets a similar fraction of the scattered signal $B_{p_r}\sim A_{p_r}$ and that all other $N-1$ input ports are in vacuum (see Fig.~\ref{fig:singlechannel}). As a practical example, let us consider the transformation for a balanced multiport beam splitter, for which the photonic amplitude operators at each independent output, $r=1,\ldots, N$, read
\begin{align}
B_{p_r}=\frac{1}{\sqrt{N}}A_{p_r}+\frac{1}{\sqrt{N}}\sum_{r'=1}^{N-1}e^{(2i\pi/N) r r'}\Upsilon_{p_r}^{r'}.\label{explicitBS}
\end{align}
Here, $\Upsilon_{p}^{r'}$ with $r'=1,\ldots, N-1$ denote the annihilation operators on the $N-1$ vacuum input ports different than channel $A$. As these extra channels are independent between each other and with channel $A$, they satisfy the commutation relations, $[\Upsilon_k^{r'},\Upsilon_{p}^{r}{}^\dag]=\delta_{rr'}\delta_{pk}$, and $[A_k,\Upsilon_{p}^{r}{}^\dag]=[U,\Upsilon_{p}^{r}]=0$, implying $[B_{p_r},B_{p_{r'}}^\dag]=\delta_{rr'}$. 

Using expression \eqref{explicitBS} in Eq.~\eqref{BcorrsFinall}, we see that the required correlations \eqref{AACorr} can be accessed by measuring at the outputs of the beam splitter. In fact, both correlations are just proportional to each other:
\begin{align}
\langle A_{p_1}\ldots A_{p_n}\rangle=N^{n/2}\langle B_{p_1}\ldots B_{p_n}\rangle.\label{CorrRelationAB}
\end{align}

Finally, we just replace $Q=A_{p_1}\ldots A_{p_n}$ into Eq.~\eqref{resultProperty} and use the relation \eqref{CorrRelationAB} to obtain the general scattering matrix elements $S_{p_1\ldots p_n k_1\ldots k_m}$ in terms of the measurable correlations $F_n(l,\vec{s})=\langle \alpha_{\vec{k}}(l,\vec{s})|U^\dag B_{p_1}\ldots B_{p_n} U|\alpha_{\vec{k}}(l,\vec{s})\rangle$, as
\begin{align}
S_{p_1\ldots p_n k_1\ldots k_m}\!=\!\frac{\sqrt{N^n}e^{|\alpha|^2}}{2^{M-1}R}\sum_{l=1}^{R}\sum_{\vec{s}}\frac{F_n(l,\vec{s})}{\prod_{j=1}^m\alpha_{k_j}^{l,\vec{s}}}+\varepsilon_{m}^{(1)}.\label{finalSupp}
\end{align}
This is the main result of the paper and is shown in Eqs.~\eqref{generalResult} and \eqref{elasticFormula} of the main text, for the general ($R=2M$) and elastic ($R=1$) cases, respectively.

The truncation error $\varepsilon_{m}^{(1)}$ appearing in the final result \eqref{finalSupp} has a closed analytical upper bound in the case that the scattering matrix $U$ conserves the number of photons (Protocol 2 of the main text). From Eqs.~\eqref{secondError}-\eqref{secondErrorCoeffs} we see that this problem is equivalent to find a closed bound for the coefficients $|h_{vm}(\vec{s})|$. Therefore, we replace $Q=A_{p_1}\ldots A_{p_m}$ in Eq.~\eqref{generalbound} and use Eq.~\eqref{explicitBS}, to re-express the bound in terms of the $B_{p_r}$ operators as
\begin{align}
|h_{vm}(\vec{s})|\!\leq\sum_{q=q_{\rm min}}^{q_{\rm max}}\!\!\frac{|\langle v-q\frac{R}{2}|U^\dag B_{p_1}\!\ldots\! B_{p_m} U|v+m+q\frac{R}{2}\rangle|}{N^{-m/2}\sqrt{(v+m+q\frac{R}{2})!(v-q\frac{R}{2})!}}.\label{generalbound2}
\end{align}
Since $U$ conserves the total number of photons, we can set $q=0$ and $U|v\rangle=e^{i\varphi_v}|v\rangle$ in Eq.~\eqref{generalbound2}, and additionally using the Cauchy-Schwarz inequality, we obtain
\begin{align}
|h_{vm}(\vec{s})|\leq\frac{|\langle v+m|(B^\dag_{p_1}B_{p_1})\ldots (B^\dag_{p_m}B_{p_m})|v+m\rangle|^{1/2}}{N^{-m/2}\sqrt{(v+m)!v!}}.\label{generalbound3}
\end{align}
The next step is to decompose the total number state $|v+m\rangle$ into components along all $N$ channels of the interferometer, and using the fact the the sum of photons in all channels is fixed, $N=\sum_{r=1}^NB_{p_r}^\dag B_{p_r}$, we can bound the numerator in Eq.~\eqref{generalbound3} by $\sqrt{(v+m)!/v!}$. Using this result in Eq.~\eqref{generalbound3} we obtain a simple bound for the $h_{vm}$ coefficients similar to Eq.~\eqref{boundh} as $|h_{vm}(\vec{s})|\leq N^{m/2}/v!$, and replacing this into Eq.~\eqref{secondErrorCoeffs}, we find the bound for the $f_{vm}$ coefficients as well, 
\begin{align}
|f_{vm}|\leq \frac{1}{v!}\frac{N^{m/2}}{\prod_{j=1}^m|\xi_{j}|}. 
\end{align}
For simplicity, let us consider the case that we prepare the multimode coherent state inputs with equal projections, $|\xi_j|=|\xi|$. The normalization condition, stated below Eq.~\eqref{SuperInput}, implies $1/|\xi|=\sqrt{\sum_{j,v=1}^M e^{-(k_j-k_{v})^2/(2\sigma)^2}}\leq M$, in the case of gaussian wave-packets as in the main text. Using this and $N=M$, we find a strict bound for the coefficients that is easy to evaluate,
\begin{align}
|f_{vm}|\leq \frac{M^{3m/2}}{v!},
\end{align}
and replacing this into Eq.~\eqref{secondError}, the strict bound for the error reads,
\begin{align}
|\varepsilon_m^{(1)}|\leq M^{3m/2}(e^{|\alpha|^2}-1).\label{simpleErrorelastic1}
\end{align}

\section{II.~Arbitrary reconstruction error by combining multiple estimates}

An intrinsic error that limits the precision of our tomography protocol is the \emph{reconstruction error} $\varepsilon_m^{(1)}=\sum_{v=1}^{\infty}|\alpha|^{2v}f_{vm}$. In this section, we show how to reduce this error by performing the scattering tomography protocol in Eq.~\eqref{finalSupp} at different laser powers and combining the outcomes in a clever way. Importantly, this allows us to perform the tomography at non-perturbative mean photon numbers $|\alpha|^2\sim 1$, lifting the requirement of attenuated coherent state inputs.

A first order estimate $E$ of the scattering matrix $S$ is given by the outcome of our protocol in Eq.~\eqref{finalSupp} as,
\begin{align}
E(|\alpha|^2)&=\!\frac{\sqrt{N^n}e^{|\alpha|^2}}{2^{M-1}R}\sum_{l=1}^{R}\sum_{\vec{s}}\frac{F_n(l,\vec{s})}{\prod_{j=1}^m\alpha_{k_j}^{l,\vec{s}}}\label{firstOrderEstimateMEasure},
\end{align}
with an error $\varepsilon_m^{(1)}=S-E(|\alpha|^2)=\sum_{v=1}^{\infty}|\alpha|^{2v}f_{vm}$, scaling as $\varepsilon_m^{(1)}=\mathcal{O}(|\alpha|^2)$. Interestingly, if we consider another first order estimate measured at a larger power $E(b|\alpha|^2)$, with $b>1$, we can combine it with $E(|\alpha|^2)$, and cancel the error term of order $\sim |\alpha|^2$. As a result, we obtain a second order estimate,
\begin{align}
E^{(2)}(|\alpha|^2)&=\frac{bE(|\alpha|^2)-E(b|\alpha|^2)}{b-1},\label{secondorderEstimate}
\end{align}
with a higher order approximation error,
\begin{align} 
\varepsilon_m^{(2)}=S-E^{(2)}(|\alpha|^2)=\sum_{v=2}^{\infty}|\alpha|^{2v}f_{vm}^{(2)},
\end{align}
that scales as $\varepsilon_m^{(2)}=\mathcal{O}(|\alpha|^4)$. Here, the modified error coefficients read $f^{(2)}_{vm}=f_{vm}[(b-b^v)/(b-1)]$. This process can be iterated up to arbitrary order using the recursion relation,
\begin{align}
E^{(\mu)}(|\alpha|^2)&=\frac{b^{\mu-1}E^{(\mu-1)}(|\alpha|^2)-E^{(\mu-1)}(b|\alpha|^2)}{b^{\mu-1}-1},\label{recursionRel}
\end{align}
which gives the estimate of order $\mu$ from two estimates of a lower order: $E^{(\mu-1)}(|\alpha|^2)$ and $E^{(\mu-1)}(b|\alpha|^2)$. In general, a $Z$-order estimate $E^{(Z)}(|\alpha|^2)$ has an error $\varepsilon_m^{(Z)}=\mathcal{O}(|\alpha|^{2Z})$ given by
\begin{align}
\varepsilon_m^{(Z)}=S-E^{(Z)}(|\alpha|^2)=\sum_{v=Z}^{\infty}|\alpha|^{2v}f^{(Z)}_{vm},\label{zordererror}
\end{align}
where the $Z$-order coefficients can expressed in terms of $f_{vm}$ in Eq.~\eqref{secondErrorCoeffs} as    
\begin{align}
f^{(Z)}_{vm}=f_{vm}\prod_{j=1}^{Z-1}\frac{(b^{Z-j}-b^v)}{(b^{Z-j}-1)}.\label{zordererrorcoeff}
\end{align}

The final step is to express a general $Z$-order estimate $E^{(Z)}(|\alpha|^2)$ in terms of first order estimates only, as these are the ones that we measure in practice via \eqref{firstOrderEstimateMEasure}. Using $Z-1$ times the recursion relation \eqref{recursionRel}, it is straightforward to check that $E^{(Z)}(|\alpha|^2)$ can always be expressed as a linear combination of $Z$ different first order estimates $E(x_q)$, measured at increasing laser powers $x_q=b^{q-1}|\alpha|^2$, with $q=1,\ldots, Z$ and $b>1$, as
\begin{align}
E^{(Z)}(|\alpha|^2)=\sum_{q=1}^{Z}w_q^{(Z)} E(x_q).\label{ZorderExpansion}
\end{align}
Here, the $Z$ coefficients $w_q^{(Z)}$ are determined recursively from Eqs.~\eqref{secondorderEstimate} and \eqref{recursionRel}. As an example, for $Z=1,\ldots, 4$, the coefficients are explicitly given by
\begin{align}
\omega^{(1)}_q&=1,\\
\omega^{(2)}_q&=\frac{\lbrace b,-1\rbrace}{b-1},\\
\omega^{(3)}_q&=\frac{\lbrace b^3,-b^2-b,1\rbrace}{(b^2-1)(b-1)},\\
\omega^{(4)}_q&=\frac{\lbrace b^6,-b^5-b^4-b^3,b^3+b^2+b,-1\rbrace}{(b^3-1)(b^2-1)(b-1)}.
\end{align}

\begin{figure}[t]
\center
\includegraphics[width=0.5\textwidth]{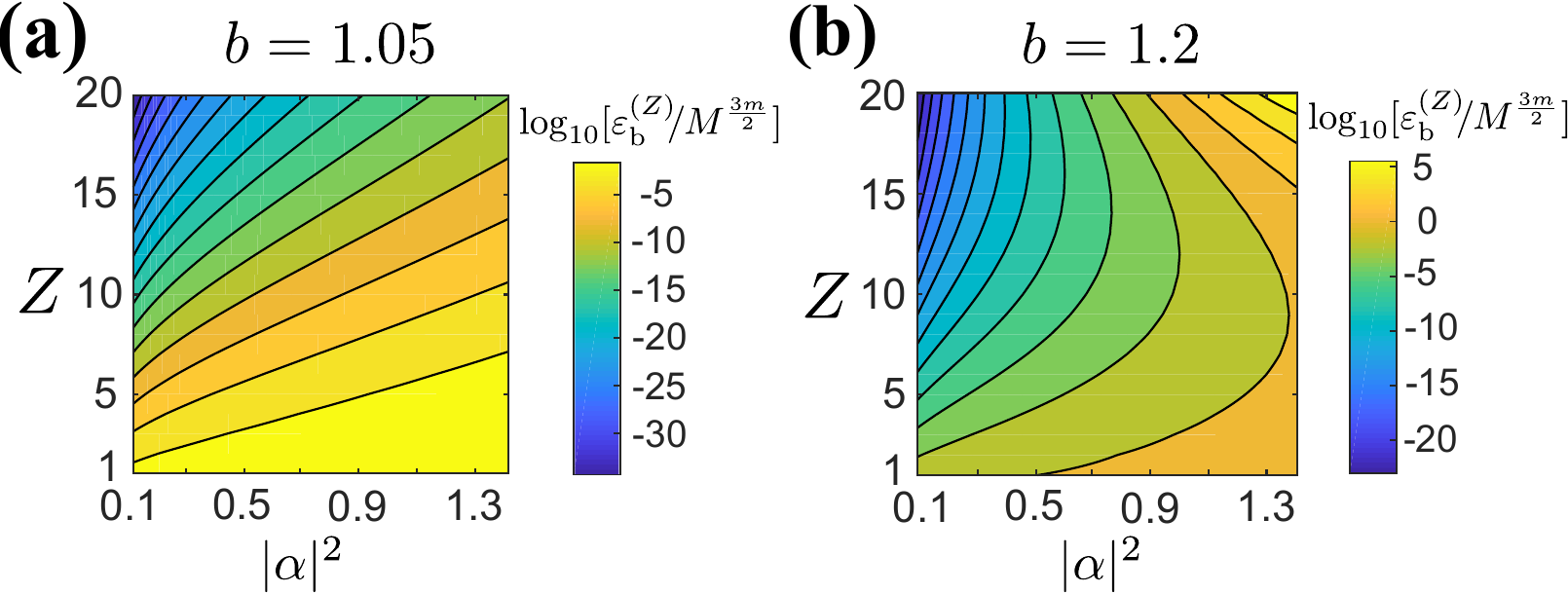}
\caption{Error bound of the $Z$-order error $\varepsilon^{(Z)}_{\rm b}/M^{3m/2}$ as a function of the smallest mean photon number $|\alpha|^2$ and the number of estimates $Z$. The estimates are obtained at increasing laser powers $x_q=b^{q-1}|\alpha|^2$ ($q=1,\dots, Z$), for $b=1.05$ in panel (a) and $b=1.2$ in panel (b). \label{fig:error}}
\end{figure}

In summary, the elements of the scattering matrix $S$ can be estimated with a $Z$-order error $\varepsilon_m^{(Z)}=\mathcal{O}(|\alpha|^{2Z})$ given in Eqs.~\eqref{zordererror}-\eqref{zordererrorcoeff} by combining $Z$ first order estimates \eqref{firstOrderEstimateMEasure}, obtained at incresing laser powers as
\begin{align}
&S_{p_1\ldots p_n k_1\ldots k_m}=\sum_{q=1}^{Z}w_q^{(Z)} E(b^{q-1}|\alpha|^2)+\varepsilon^{(Z)}_m.
\end{align}
Importantly, the $Z$-order error $\varepsilon^{(Z)}_m=\mathcal{O}(|\alpha|^{2Z})$ is strongly suppressed by its prefactor, which allows one to perform the scattering tomography protocol at non-perturbative laser powers $|\alpha|^2\gtrsim 1$. We show this explicitly by numerically evaluating the upper bound of the $Z$-order error $\varepsilon_{\rm b}^{(Z)}$ in the case of elastic scattering $m=n$, which is obtained from Eqs.~\eqref{simpleErrorelastic1}, \eqref{zordererror}, and \eqref{zordererrorcoeff} as 
\begin{align}
|\varepsilon_m^{(Z)}|\leq \varepsilon_{\rm b}^{(Z)}=M^{3m/2}\sum_{v=Z}^{\infty}\frac{|\alpha|^{2v}}{v!}\prod_{j=1}^{Z-1}\frac{(b^{Z-j}-b^v)}{(b^{Z-j}-1)}.
\end{align}
In Fig.~\ref{fig:error}a-b, we plot $\varepsilon^{(Z)}_{\rm b}/M^{3m/2}$ in logarithmic scale, as a function of the minimal mean photon number $|\alpha|^2$ and the number of estimates $Z$, for for two values of the factor $b$. For $b=1.05$, we indeed get a good error bound of $\varepsilon^{(Z)}_{\rm b}/M^{3m/2}\sim 10^{-5}$ for $|\alpha|^2\sim 1$ and a moderate number of estimates $Z\sim 10$. Figure \ref{fig:deconvolution}a of the main text shows the same bound $\varepsilon^{(Z)}_{\rm b}$ and the same parameters, but evaluated in the specific case $M=m=2$.

In general, the error suppression is stronger for smaller $b$, as it becomes clear by comparing Figs.~\ref{fig:error}a and \ref{fig:error}b. Experimentally, the value of this factor $b$ is restricted by the precision to increase the laser power as $x_q=b^{q-1}|\alpha|^2$. Finally, notice that for a given starting mean photon number $|\alpha|^2$ and a fixed value of the $b$ factor, there will be an optimal number of estimates $Z$ that gives the maximum supression of the error. For $Z$ higher than the optimal, the error $\varepsilon^{(Z)}_m$ will actually increase due to its prefactors (see Fig.~\ref{fig:error}b).

\section{III.~Deconvolution with gaussian wave packets in dispersive channels}

To be sensitive to nonlinear multiphoton scattering events, our scattering protocol requires wave packet input modes in the case the quantum number $k_j$ describes a continuous degree of freedom such as momentum or frequency. Therefore, in the following analysis we want to distinguish between continuous and discrete photon degrees of freedom, which we label by $k_j$ and $\lambda_j$, respectively. For instance, photon polarization and/or path can be included in the $\lambda_j$ quantum numbers. Using this separation of labels, we assume that the generic bosonic modes in Eq.~\eqref{SuperInput} are prepared as  
\begin{align}
A_{k_j}^{\lambda_j}=\int \mathrm{d}k' \psi_{k_j}^{\ast}(k')a_{k'}^{\lambda_j},\label{wavepacketSM}
\end{align}
where $\psi_{k_j}(k')$ is the square-normalized wave packet profile for the continuous variable $k'$, centered at $k_j$, and $a_{k'}^{\lambda_j}$ describe monochromatic photons with quantum numbers $k'$ and $\lambda_j$, satisfying standard communation relations $[a_{k}^{\lambda_{j}},(a_{k'}^{\lambda_{j}'})^\dag]=\delta_{\lambda_j\lambda_{j'}}\delta(k-k')$.

The choice of input modes in Eq.~\eqref{wavepacketSM} implies that our scattering protocol gives us direct access to the `wave packet' scattering matrix elements only, given in general by 
\begin{align}
S_{p_1\ldots p_n, k_1\ldots k_m}^{\lambda_1'\ldots\lambda_n',\lambda_1\ldots\lambda_m}=\langle 0|A_{p_1}^{\lambda_1'}\ldots A_{p_n}^{\lambda_n'} U (A_{k_1}^{\lambda_1})^\dag\ldots (A_{k_m}^{\lambda_m})^\dag|0\rangle.
\end{align}
As explained in the main text, the scattering matrix elements in the monochromatic Fock basis, $\Smono_{p_1\ldots p_n k_1\ldots k_m}^{\lambda_1'\ldots\lambda_n',\lambda_1\ldots\lambda_m}=\langle 0|a_{p_1}^{\lambda_1'}\ldots a_{p_n}^{\lambda_n'} U (a_{k_1}^{\lambda_1})^\dag\ldots (a_{k_m}^{\lambda_m})^\dag|0\rangle$, can still be accessed by using Eq.~\eqref{wavepacketSM} and inverting the integral relation,
\begin{align}
\label{relationSDIscrete}
S_{p_1\ldots p_n, k_1\ldots k_m}^{\lambda_1'\ldots\lambda_n',\lambda_1\ldots\lambda_m}&=\int\!\ldots\!\int\mathrm{d}^{m}k'\mathrm{d}^{n}p'\ \Smono_{p_1\ldots p_n k_1\ldots k_m}^{\lambda_1'\ldots\lambda_n',\lambda_1\ldots\lambda_m}\times\notag\\
&\times \prod_{j=1}^m \psi_{k_j}(k_j')\prod_{r=1}^n \psi^{\ast}_{p_r}(p_r').
\end{align}
Therefore, one needs to perform a deconvolution on the continuous variables $k_1,\ldots, k_m,p_1,\ldots, p_n$ for each combination of the discrete variables $\lambda_1,\ldots,\lambda_m,\lambda_1',\ldots,\lambda_n'$, which in turn requires the measurement of the wave packet matrix elements for all convinations of quantum numbers $S_{p_1\ldots p_n, k_1\ldots k_m}^{\lambda_1'\ldots\lambda_n',\lambda_1\ldots\lambda_m}$. As we are mainly interested in illustrating how to perform a single deconvolution, in the remainder of the section, as well as in the main text, we assume that all discrete quantum numbers are equal and we absorb them in the definition of the photon operators $A_{k_j}$ and $a_{k_j}$. This simplifies the equations considerably, and we just have to keep in mind that to describe additional internal degrees of freedom of photons, we have to repeat the same deconvolution procedure for each combination of of the discrete labels $\lambda_1,\ldots,\lambda_m,\lambda_1',\ldots,\lambda_n'$.

In this section, we show how to analytically solve the deconvolution integral for the monochromatic single- and two-photon scattering matrices in Eq.~\eqref{scatteringStructure} of the main text, which corresponds to the more general Eq.~\eqref{relationSDIscrete} in the case the continuous degree of freedom $k_j$ corresponds to momentum or frequency, and all photons have the same discrete degrees of freedom. To do so, we assume wave-packets with gaussian shape and a possibly dispersive photonic channel, as shown below. 

In Sec.~III.C, we show the analytical expressions for the single- and two-photon scattering matrices used in Figs.~\ref{fig:deconvolution}(b)-(d) of the main text to test the gaussian deconvolution formulas.

\subsection{A.~Gaussian deconvolution in a dispersive channel}

As we want to discuss photons in a dispersive channel, the deconvolution integrals turn out to be simpler when working with frequency modes instead of momentum. Therefore, in this subsection we derive all deconvolution formulas in terms of frequency-direction modes, and in the next subsection we map the results to momentum, connecting to the specific formulas shown in the main text. 

In the special case that the dispersion relation is symmetric, $\omega_{k}=\omega_{-k}$, and the group velocity is antisymmetric, $v_k=-v_{-k}$, the propagating momentum modes $a_k$ can be unambiguously related to frequency-direction modes $a^{\rho}_\omega$, as
\begin{align}
a^{\rho}_\omega=\frac{a_{\rho|k|}}{\sqrt{|v_{k}|}}.\label{FreqMomRel}
\end{align}
Here, the mapping assumes that photons with $k>0$ propagate to the right ($v_{k}>0$), while photons with $k<0$ propagate to the left ($v_{k}<0$), and we identify these directions with the quantum number $\rho={\rm sign}(k)=\pm$. In addition, the frequency $\omega$ of the photons has a one-to-one correspondence with the wavenumber $|k|$ via the invertible function $\omega=\omega_{|k|}$. Note that this relation \eqref{FreqMomRel} is not valid for $k=0$, since the group velocity necessarily vanishes $v_0=0$. Nevertheless, we will always consider propagating wave packets \eqref{wavepacketSM} with negligible components on this $k=0$ mode. 

The wave packet operators in Eq.~\eqref{wavepacketSM} can be expressed in terms of frequency-direction modes as 
\begin{align}
A_k^\dag=A_{\rho|k|}^\dag=A^{\rho}_\omega{}^\dag=\sum_{\rho'}\int d\omega' \tilde{\psi}^{\rho\rho'}_{\omega}(\omega') a_{\omega'}^{\rho'}{}^\dag,
\end{align}
where the corresponding wave packet profile is related to its momentum counterpart by
\begin{align}
\tilde{\psi}^{\rho\rho'}_\omega(\omega')=\frac{\psi_{k}(k')}{\sqrt{|v_{k'}|}}.
\end{align}
In particular, we choose wave packets with a gaussian profile defined as
\begin{align}
\psi^{\rho\rho'}_{\omega}(\omega')={}&\sqrt{G_{\tilde{\sigma}}(\omega-\omega')}\delta_{\rho\rho'},\, {\rm with}\\
G_{\tilde{\sigma}}(\omega')={}&(\pi\tilde{\sigma}^2)^{-1/2}e^{-(\omega'/\tilde{\sigma})^2},\label{GaussianFrequency}
\end{align}
such that the wave packet mode $A_\omega^{\rho}{}^\dag=\int d\omega' G_{\tilde{\sigma}}(\omega'-\omega)^{1/2}a^{\rho}_{\omega'}{}^\dag$ is built exclusively from monochromatic frequency modes $a^{\rho}_{\omega'}$ propagating in the same direction $\rho$. We also assume that the central frequency of the wave packet $\omega$ is much larger than its width, $\tilde{\sigma}\ll\omega$, to ensure a vanishing component on the non-propagating mode $\omega=0$.

Using all these considerations, we can re-express the general integral equation \eqref{relationS} of the main text, in terms of frequency-direction modes, as
\begin{align}
\label{relationSfrequency}
{S}_{\omega_1\ldots \omega_n\nu_1\ldots \nu_m}^{\rho_1\ldots\rho_n \eta_1\ldots\eta_m}&=\int\!\ldots\!\int\mathrm{d}^{m}\nu'\mathrm{d}^{n}\omega'\ \Smono_{\omega_1'\ldots \omega_n'\nu_1'\ldots \nu_m'}^{\rho_1\ldots\rho_n \eta_1\ldots\eta_m}\times\\&\times \prod_{j=1}^m G_{\tilde{\sigma}}(\nu_j'-\nu_j)^{1/2}\prod_{r=1}^n G_{\tilde{\sigma}}(\omega_r'-\omega_r)^{1/2},\notag
\end{align}
where $\Smono_{\omega_1\ldots \omega_n\nu_1\ldots \nu_m}^{\rho_1\ldots\rho_n \eta_1\ldots\eta_m}=\langle 0|a_{\omega_1}^{\rho_1}\ldots a_{\omega_n}^{\rho_n} U a_{\nu_1}^{\eta_1}{}^\dag\ldots a_{\nu_m}^{\eta_m}{}^\dag|0\rangle$ denote the monochromatic scattering elements from $m$ incoming photons with frequencies $\nu_1,\ldots, \nu_m$, and directions $\eta_1,\ldots, \eta_m$, to $n$ outgoing ones with frequencies $\omega_1,\ldots, \omega_n$, and directions $\rho_1,\ldots, \rho_n$, respectively.

In the following, we specialize the discussion to the single- and two-photon matrix elements of a scatterer with a single ground state. In this case, the conservation of energy allows us to decompose the monochromatic scattering components as in Eq.~\eqref{scatteringStructure} of the main text:
\begin{align}
\Smono_{\omega_1\nu_1}^{\rho_1\eta_1}={}&\chi^{\rho_1\eta_1}_{\omega_1}\delta(\omega_1-\nu_1),\label{monofreq1}\\
\Smono_{\omega_1\omega_2\nu_1\nu_2}^{\rho_1\rho_2\eta_1\eta_2}={}&\Smono_{\omega_1\nu_1}^{\rho_1\eta_1}\Smono_{\omega_2\nu_2}^{\rho_2\eta_2}+\Smono_{\omega_1\nu_2}^{\rho_1\eta_2}\Smono_{\omega_2\nu_1}^{\rho_2\eta_1}\label{monofreq2}\\
{}&+i\Tmono_{\omega_1\omega_2\nu_1\nu_2}^{\rho_1\rho_2\eta_1\eta_2}\delta(\omega_1+\omega_2-\nu_1-\nu_2).\notag
\end{align}
Here, $\chi_{\omega}^{\rho\eta}$ contain the reflection $r_\omega=\chi_{\omega}^{-\rho,\rho}$ and transmission $t_\omega=\chi_{\omega}^{\rho,\rho}$ coefficients, and the nonlinear contribution $\Tmono_{\omega_1\omega_2\nu_1\nu_2}^{\rho_1\rho_2\eta_1\eta_2}$ describes photon-photon interactions mediated by the scatterer. 

Using Eqs.~\eqref{monofreq1}-\eqref{monofreq2} in Eq.~\eqref{relationSfrequency}, simplifies the integral relations by reducing their dimensionality. In particular, the single-photon scattering matrix satisfies,
\begin{align}
S_{\omega_1\nu_1}^{\rho_1\eta_1}=\! e^{-\frac{(\omega_1-\nu_1)^2}{4\tilde{\sigma}^2}}\!\!\int\! d\omega_1'\ \chi_{\omega_1'}^{\rho_1\eta_1} G_{\tilde{\sigma}}(\omega_1'-\frac{[\omega_1+\nu_1]}{2}).\label{singleConvolutionSM}
\end{align}
On the other hand, the two-photon `wave packet' scattering matrix can be also decomposed into linear and nonlinear contributions,
\begin{align}
S_{\omega_1\omega_2\nu_1\nu_2}^{\rho_1\rho_2\eta_1\eta_2}\!=S_{\omega_1\nu_1}^{\rho_1\eta_1}S_{\omega_2\nu_2}^{\rho_2\eta_2}+S_{\omega_1\nu_2}^{\rho_1\eta_2}S_{\omega_2\nu_1}^{\rho_2\eta_1}+iT_{\omega_1\omega_2\nu_1\nu_2}^{\rho_1\rho_2\eta_1\eta_2},
\end{align}
where the non-linear part reads,
\begin{align}
{}&T_{\omega_1\omega_2\nu_1\nu_2}^{\rho_1\rho_2\eta_1\eta_2}=2\tilde{\sigma}\sqrt{\pi}e^{-\frac{(\bar{\omega}-\bar{\nu})^2}{2\tilde{\sigma}^2}}\!\!\int\! d\bar{\omega}' d\Delta_{\omega}' d\Delta_{\nu}' \mathcal{W}(\bar{\omega}',\Delta_{\omega}',\Delta_{\nu}')\notag\\
{}&\qquad\qquad\qquad\qquad\times\Tmono_{\bar{\omega}'-\Delta_\omega',\bar{\omega}'+\Delta_\omega',\bar{\omega}'-\Delta_\nu',\bar{\omega}'+\Delta_\nu'}^{\rho_1\rho_2\eta_1\eta_2}.\label{convolutionTwoPhotonSM}
\end{align}
Here, the kernel $\mathcal{W}$ is the product of three Gaussians, 
\begin{align}
\mathcal{W}=G_{\tilde{\sigma}}(\sqrt{2}[\bar{\omega}'-\frac{(\bar{\omega}+\bar{\nu})}{2}])G_{\tilde{\sigma}}(\Delta_\omega'-\Delta_\omega)G_{\tilde{\sigma}}(\Delta_\nu'-\Delta_\nu),
\end{align}
and we conveniently defined the new variables $\bar{\omega}=(\omega_1+\omega_2)/2$, and $\bar{\nu}=(\nu_1+\nu_2)/2$, $\Delta_\nu=(\nu_2-\nu_1)/2$, and $\Delta_\omega=(\omega_2-\omega_1)/2$.

Interestingly, both Eqs.~\eqref{singleConvolutionSM} and \eqref{convolutionTwoPhotonSM} can be analytically inverted as they involve a product of Gaussian kernels for independent integration variables. Therefore, following Refs.~\cite{ulmer2003,ulmer2010,fang1994} of the main text, we obtain
\begin{align}
{}&\chi_{\omega_1'}^{\rho_1\eta_1}=\int\! d\omega_1\  \mathcal{K}_{\tilde{\sigma}}(\omega_1-\omega_1')S_{\omega_1\omega_1}^{\rho_1\eta_1},\label{singleDeConvolutionSM}\\
{}&\Tmono_{\omega_1\omega_2\nu_1\nu_2}^{\rho_1\rho_2\eta_1\eta_2}=\frac{1}{\tilde{\sigma}\sqrt{\pi}}\int\! d\bar{\omega}' d\Delta_{\omega}' d\Delta_{\nu}' T_{\bar{\omega}'-\Delta_\omega',\bar{\omega}'+\Delta_\omega',\bar{\omega}'-\Delta_\nu',\bar{\omega}'+\Delta_\nu'}^{\rho_1\rho_2\eta_1\eta_2}\notag\\
{}&\quad\times \mathcal{K}_{\tilde{\sigma}}(\sqrt{2}[\bar{\omega}'-\bar{\omega}])\mathcal{K}_{\tilde{\sigma}}(\Delta_\omega'-\Delta_\omega)\mathcal{K}_{\tilde{\sigma}}(\Delta_\nu'-\Delta_\nu),\label{DeConvolutionTwoPhotonSM}
\end{align}
where the inverse Gaussian kernel $\mathcal{K}_{\tilde{\sigma}}(\omega)$ is given by
\begin{align}
K_{\tilde{\sigma}}(\omega)=G_{\tilde{\sigma}}(\omega)\sum_{q=0}^\infty \frac{(-1)^q}{2^q q!}H_{2q}\left(\frac{\omega}{\tilde{\sigma}}\right).
\end{align}
Here, $H_{q}(x)=(-1)^q e^{x^2}\partial_x^q (e^{-x^2})$ are  Hermite polynomials and the series is convergent because 
$S_{\omega_1\omega_1}^{\rho_1\eta_1}$ and $ T_{\omega_1\omega_2\nu_1\nu_2}^{\rho_1\rho_2\eta_1\eta_2}$ decay exponentially fast.

These last Eqs.~\eqref{singleDeConvolutionSM}-\eqref{DeConvolutionTwoPhotonSM} are direct analytical relations for the single- and two-photon scattering coefficients in terms of the `wave packet' counterparts, measurable with our tomography protocol. Moreover, they are valid for any dispersive photonic channel, satisfying $\omega_{k}=\omega_{-k}$ and $v_{k}=-v_{-k}$. In the next subsection, we connect to the expressions stated in the main text, by transforming to momentum variables, and specializing to a linear dispersion relation $\omega_k=c|k|$.

\subsection{B.~Gaussian deconvolution in momentum modes for a non-dispersive channel}

The monochromatic scattering elements in momentum and frequency basis can be related using Eq.~\eqref{FreqMomRel} as 
\begin{align}
\Smono_{p_1k_1}={}&|v_{p_1}v_{k_1}|^{1/2}\Smono_{\omega_1\nu_1}^{\rho_1\eta_1},\\
\Smono_{p_1p_2k_1k_2}={}&|v_{p_1}v_{p_2}v_{k_1}v_{k_2}|^{1/2}\Smono_{\omega_1\omega_2\nu_1\nu_2}^{\rho_1\rho_2\eta_1\eta_2},
\end{align}
where we used the identifications, $\omega_j=\omega_{p_j}$, $\nu_j=\omega_{k_j}$, $\rho_j={\rm sign}(p_j)$, and $\eta_j={\rm sign}(k_j)$. In addition, using the explicit formulas \eqref{monofreq1}-\eqref{monofreq2}, we obtain the same formulas for the momentum monochromatic elements in the main text,
\begin{align}
\Smono_{p_1k_1}={}&c\chi_{p_1k_1}\delta(\omega_{p_1}-\omega_{k_1}),\\
\Smono_{p_1p_2k_1k_2}={}&\Smono_{p_1k_1}\Smono_{p_2k_2}+\Smono_{p_1k_2}\Smono_{p_2k_1}\\
{}&+ic\Tmono_{p_1p_2k_1k_2}\delta(\omega_{p_1}+\omega_{p_2}-\omega_{k_1}-\omega_{k_2}),\notag
\end{align}
but now with the coefficients given for dispersive media as,
\begin{align}
\chi_{p_1k_1}={}&\frac{1}{c}|v_{p_1}v_{k_1}|^{1/2}\chi^{\rho_1\eta_1}_{\omega_{p_1}},\\
\Tmono_{p_1p_2k_1k_2}={}&\frac{1}{c}|v_{p_1}v_{p_2}v_{k_1}v_{k_2}|^{1/2}\Tmono_{\omega_{p_1}\omega_{p_2}\omega_1\nu_2}^{\rho_1\rho_2\eta_1\eta_2}.
\end{align}
The wave packet profile used in Eq.~\eqref{GaussianFrequency} to perform the deconvolutions, can be recast in momentum as
\begin{align}
\psi_k(k')=\left[\frac{|v_{k'}|}{c}G_{\sigma}\left(\frac{\omega_{k'}-\omega_{k}}{c}\right)\right]^{1/2}\!\!\!\delta_{{\rm sign}(k),{\rm sign}(k')},
\end{align}
where the $\sigma=\tilde{\sigma}/c$ is the momentum width used in the main text. 

Finally, when evaluating the above expressions for a linear dispersion relation $\omega_k=c|k|$ and forward scattering $k_j,p_j>0$, as in the main text. In particular, the wave packet profile reduces simply to $\psi_k(k')=G_{\sigma}\left(k'-k\right)^{1/2}$ and the scattering coefficients read,
\begin{align}
&t_{p_1}=\chi^{++}_{\omega_{p_1}}=\int\! dp_1'\  \mathcal{K}_{\sigma}(p_1'-p_1)S_{p_1'p_1'},\\
&\Tmono_{p_1p_2k_1k_2}=\frac{1}{\sqrt{\pi}\sigma}\!\int\!\mathrm{d}\hat{k}'\mathrm{d}\Delta_p'\mathrm{d}\Delta_k' T_{\hat{k}'-\Delta_p',\hat{k}'+\Delta_p',\hat{k}'-\Delta_k',\hat{k}'+\Delta_k'}\notag\\
&\quad\times\mathcal{K}_{\sigma}(\sqrt{2}[\hat{k}'-\hat{k}])\mathcal{K}_{\sigma}(\Delta_p'-\Delta_p)\mathcal{K}_{\sigma}(\Delta_k'-\Delta_k),
\end{align}
where $\hat{k}=(k_1+k_2)/2=(p_1+p_2)/2$, $\Delta_p=(p_2-p_1)/2$, and $\Delta_k=(k_2-k_1)/2$.

\subsection{C.~One- and two-photon scattering matrix for a qubit weakly coupled to a 1D waveguide}

In this subsection, we explicitly state the results shown in Refs.~\cite{shen07,fan10} of the main text, for the one- and two-photon scattering matrices of a two-level system weakly coupled to a 1D photonic waveguide. We use these analytical results in Figs.~\ref{fig:deconvolution}(b)-(d) of the main text to test the deconvolution formulas shown above. In particular, the monochromatic single-photon scattering matrix has the form of Eq.~(\ref{monofreq1}) with the reflection $r_\omega=\chi_{\omega}^{-\rho,\rho}$ and transmission $t_\omega=\chi_{\omega}^{\rho,\rho}$ coefficients given by
\begin{align}
r_{\omega}=\frac{-1}{1-i(\omega-\omega_0)/\gamma},\qquad t_\omega=1+r_\omega.
\end{align}
Here $\omega_0$ is the transition frequency of the two-level system and $\gamma$ its decay rate. The two-photon scattering matrix is of the form of Eq.~(\ref{monofreq2}), with the two-photon interaction strength given by
\begin{align}
\Tmono_{\omega_1\omega_2\nu_1\nu_2}^{\rho_1\rho_2\eta_1\eta_2}=-\frac{i}{\pi\gamma}r_{\omega_1}r_{\omega_2}(r_{\nu_1}+r_{\nu_2}).\label{twophotonanalytic}
\end{align}
This is the nonlinearity we replace in Eq.~(\ref{convolutionTwoPhoton}) of the main text (with the proper change of variables as explained in the previous subsections) to evaluate the prediction for a scattering experiment with gaussian wave packets and a weakly coupled qubit. We then perform a deconvolution of the result using Eq.~(\ref{inverseTwoPhoton}) of the main text, and check that the result agrees with the initial Eq.~(\ref{twophotonanalytic}), which allows us to test the deconvolution formulas.

\section{IV.~Imperfections in the input state preparation}

In this section we discuss errors in the reconstruction of the scattering matrix elements, which are caused by imperfections in the preparation of the coherent state inputs. The analysis includes deviations in the relative phases $s_j=\pm 1+\delta s_j^{(\pm)}$ [cf.~Sec.~IV.A], laser power $|\alpha|^2+\delta n$ [cf.~Sec.~IV.B], and global phases $\phi_l=\pi l/M+\delta \phi_l$ [cf.~Sec.~IV.C].

\subsection{A.~Deviations in relative phases}

In the case the scatterer conserves the photon number, we can apply Protocol 2 of the main text, and therefore the preparation of the various input states requires only the control of relative phases $s_j=\pm 1$, for $j=1,\ldots, M$. If we go back to Eq.~(\ref{IntermediateL}) in the derivation of the protocol and assume a small deviation, $s_j=\pm 1+\delta s_j^{(\pm)}$, in the ideal relative phases, we obtain the relation, 
\begin{align}
\langle 0|QUA^\dag_{k_1}\!\ldots \!A^\dag_{k_m}|0\rangle\!=\!\sum_{\vec{s}}\frac{C_{0m}(\vec{s})}{2^{M-1}\xi_1^{(1)}\xi_2^{(s_2)}\!\!\ldots \xi_m^{(s_m)}}+\varepsilon_{\rm sign},\label{signerror}
\end{align}
where the `sign' error $\varepsilon_{\rm sign}$ is given to first order in the small deviations $|\delta s_j^{(\pm)}|\ll 1$ as
\begin{align}
\varepsilon_{\rm sign}&=\sum_{l=2}^m \frac{|\xi_1|}{2|\xi_l|}\delta \bar{s}_l \langle 0|QU (A_{k_1}^\dag)^2\!\!\!\prod_{j=2,[j\neq l]}^m\!\! A_{k_j}^\dag|0\rangle\\
&-\sum_{l=2}^m \frac{|\xi_1|}{2|\xi_l|}\delta \bar{s}_l \langle 0|QU (A_{k_l}^\dag)^2\!\!\!\prod_{j=2,[j\neq l]}^m\!\! A_{k_j}^\dag|0\rangle\notag\\
&+\sum_{l=2}^m \sum_{q=2}^m\frac{|\xi_q|^2}{6|\xi_1||\xi_l|}\delta \bar{s}_l \langle 0|QU (A_{k_q}^\dag)^3\!\!\!\prod_{j=2,[j\neq l,q]}^m\!\! A_{k_j}^\dag|0\rangle\notag\\
&+\sum_{l=2}^m \sum_{q=m+1}^M\frac{|\xi_q|^2}{2|\xi_1||\xi_l|}\delta \bar{s}_l \langle 0|QU (A_{k_q}^\dag)^2\!\!\!\prod_{j=2,[j\neq l]}^m\!\! A_{k_j}^\dag|0\rangle\notag\\
&-\sum_{l=m+1}^M \sum_{q=m+1}^M\frac{|\xi_l|}{|\xi_1|}\delta \bar{s}_l \langle 0|QU A_{k_l}^\dag\prod_{j=2}^m\!\! A_{k_j}^\dag|0\rangle+\mathcal{O}(|\delta s_j^{(\pm)}|^2),\notag
\end{align}
with $\delta \bar{s}_j$ he average deviation $\delta \bar{s}_j=(\delta s_j^{(+)}+\delta s_j^{(-)})/2$.

The rest of the derivation for the reconstruction formula is the same as in Sec.~I.B. In particular, when replacing Eq.~(\ref{finalSingle}) into the imperfect relation (\ref{signerror}), we obtain the usual estimate of the scattering matrix in Eqs.~(\ref{resultProperty}) and (\ref{firstOrderEstimateMEasure}), but now with the extra 'sign' error given above,
\begin{align}
S=E(|\alpha|^2)+\varepsilon_m^{(1)}+\varepsilon_{\rm sign}.
\end{align}
Notice that this small error in the relative phases is linear in the deviations, and thus can be controlled provided $\varepsilon_{\rm sign}\sim |\delta s_j^{(\pm)}|\ll 1$. Since $\varepsilon_{\rm sign}$ does not depend on the laser power $|\alpha|^2$, the formula for the $Z$-order estimate is straightforwardly derived by reapeating the procedure in Sec.~II., obtaining simply
\begin{align}
S=E^{(Z)}(|\alpha|^2)+\varepsilon_m^{(Z)}+\varepsilon_{\rm sign}.\label{estimationSignerror}
\end{align}
with $E^{(Z)}(|\alpha|^2)$ and $\varepsilon_m^{(Z)}$ as given in Eqs.~(\ref{ZorderExpansion}) and (\ref{zordererror}).

\subsection{B.~Imperfections in laser power input}

In the case there are imperfections in setting the mean photon number of the input states, $|\alpha|^2+\delta n$, the procedure in Sec.~II for combining $Z$ estimates at well defined laser powers, $x_q=b^{q-1}|\alpha|^2$, will be affected by an error as well. For a first order estimate, this error appears as
\begin{align}
&E(|\alpha|^2+\delta n)=E(|\alpha|^2)+\varepsilon_{\alpha}^{(1)},\quad\rm{with}\\
&\varepsilon_{\alpha}^{(1)}(|\alpha|^2)=-\sum_{v=1}^{\infty}|\alpha|^{2v}f_{vm}\left[\left(1+\frac{\delta n}{|\alpha|^2}\right)^v-1\right].
\end{align}
Using the above expression in the procedure of Sec.~II to combine $Z$ estimates, we generalize Eq.~(\ref{estimationSignerror}),
\begin{align}
S=E^{(Z)}(|\alpha|^2)+\varepsilon_m^{(Z)}+\varepsilon_{\rm sign}+\varepsilon_{\alpha}^{(Z)},\label{estimatePowerError}
\end{align}
with the extra error $\varepsilon_{\alpha}^{(Z)}$ due to imperfections in the laser power, given by
\begin{align}
\varepsilon_{\alpha}^{(Z)}(|\alpha|^2)=-\sum_{v=1}^{\infty}|\alpha|^{2v}f_{vm}^{(Z)}\left[\left(1+\frac{\delta n}{|\alpha|^2}\right)^v-1\right].
\end{align}
Notice that for $\delta n/|\alpha|^2\ll 1$ this error is also linear in the deviations,
\begin{align}
\varepsilon^{(Z)}_{\alpha}(|\alpha|^2)=-\frac{\delta n}{|\alpha|^2}\sum_{v=1}^{\infty}|\alpha|^{2v}vf_{vm}^{(Z)}+\mathcal{O}\left(\left[\frac{\delta n}{|\alpha|^2}\right]^2\right),
\end{align}
and it is small $|\varepsilon^{(Z)}_{\alpha}|\sim \delta n/|\alpha|^2\ll 1$ since high order terms in the sum (with $v\gg 1$) are strongly suppressed by the coefficients $|f_{vm}^{(Z)}|\sim 1/v!$ in Eq.~(\ref{zordererrorcoeff}).

\subsection{C.~Deviations in global phases}

In this last section we discuss errors due to deviations in the global phase of the input states, $\phi_l=\pi l/M+\delta \phi_l$, for $l=1,\ldots, 2M$. This is needed in the case that we do not know \emph{a priori} if the scatterer conserves the photon number, and we have to apply the general Protocol 1 of the main text. If we go back to Eq.~(\ref{finalSingle}) in the derivation of the protocol (with $\lambda\!=\!1$), the deviations $\delta\phi_l$ will add an extra error to this expression, obtaining 
\begin{align}
C_{0m}(\vec{s})=\frac{e^{|\alpha|^2}}{2M}\sum_{l=1}^{2M}\frac{\langle Q\rangle(l,\vec{s})}{[\alpha(l)]^m}+\epsilon_m^{(1)}+\epsilon_m^{\phi(1)}.\label{C0mError}
\end{align}
In addition to the usual approximation error $\epsilon_m^{(1)}$, the error due to imperfect global phases $\delta\phi_l$ reads,
\begin{align}
&\epsilon_m^{\phi(1)}=-\sum_{u=1}^{\infty}|\alpha|^{u-m}\!\!\!\varsum_{d=-u+2}^u\!\! T_{ud}\ \delta y^{\phi}_{dm},\quad \rm{with}\\
&\delta y^\phi_{dm}=\frac{1}{2M}\sum_{l=1}^{2M}e^{i \frac{\pi l}{M}(d-m)}(e^{i\delta \phi_l d}-1).
\end{align}

Including the error due to imperfect relative phases, we replace Eq.~(\ref{C0mError}) into Eq.~(\ref{signerror}), and following the usual derivation, we obtain a more general expression for the first order estimate of $S$,
\begin{align}
S=E(|\alpha|^2)+\varepsilon_m^{(1)}+\varepsilon_{\rm sign}+\varepsilon_m^{\phi(1)}.
\end{align}
Here, $\varepsilon_m^{(1)}$ and $\varepsilon_{\rm sign}$ are the approximation and sign errors already discussed, and the first order error due to imperfections in the global phases reads
\begin{align}
&\varepsilon_m^{\phi(1)}=\sum_{u=1}^{\infty}|\alpha|^{u-m}t_{um},\quad {\rm with}\\
&t_{um}=-\sum_{\vec{s}}\varsum_{d=-u+2}^{u}\frac{T_{ud}(\vec{s})\delta y^{\phi}_{dm}}{2^{M-1}\prod_{j=1}^m\xi_j^{(s_j)}}.
\end{align}

Additionally, if we now include the error due to imperfections in the laser power as in Eq.~(\ref{estimatePowerError}), with $\delta n/|\alpha|^2\ll 1$, and combine $Z$ estimates as in Sec.~II, we derive the most general $Z$-order estimate for $S$,
\begin{align}
S=E^{(Z)}(|\alpha|^2)+\varepsilon_m^{(Z)}+\varepsilon_{\rm sign}+\varepsilon_{\alpha}^{(Z)}+\varepsilon_m^{\phi(Z)},
\end{align}
with the $Z$-order error due to imperfect global phases given by 
\begin{align}
\varepsilon_m^{\phi(Z)}=\sum_{u=1}^{\infty}|\alpha|^{u-m}t_{um}\prod_{j=1}^{Z-1}\frac{(b^{Z-j}-b^{(u-m)/2})}{(b^{Z-j}-1)}.
\end{align}
Here, only the few first terms in the expansion contribute to the error as $|T_{ud}|\sim 1/u!$ in $t_{um}$ strongly suppresses higher order terms with $u\gg 1$. Finally, notice that if we use moderate laser powers $|\alpha|\gtrsim 1$ and assume small deviations in the global phases $|\delta \phi_l|\ll 1$, the resulting error is small and linear in the deviations, scaling approximately as $|\varepsilon_m^{\phi,(Z)}|\sim |\delta \phi_l|/|\alpha|^{m-1}\sim |\delta \phi_l| \ll 1$.

\end{document}